\documentclass[aps,prd,twocolumn,twoside,superscriptaddress,floatfix,nofootinbib]{revtex4-1}

\usepackage{graphicx}
\usepackage{ifthen,amsmath,amssymb,bm}
\usepackage[dvipsnames]{xcolor}
\usepackage{hyperref}
\usepackage{float}
\usepackage{aas_macros}
\usepackage[export]{adjustbox}
\usepackage{bookmark}

\newcommand{\vk}{\mathbf{k}}
\newcommand{\nhat}{\hat{\mathbf{n}}}

\newcommand{\mpc}{h^{-1}\mathrm{Mpc}}

\newcommand{\msol}{h^{-1}M_{\odot}}

\newcommand{\HEALPIX}{{\textsc{HEALPix}}}

\newcommand{\lmax}{\ell_{\mathrm{max}}}
\newcommand{\fsky}{f_{\mathrm{sky}}}

\newcommand{\Cme}{\mathcal{C}_\ell^{me}}
\newcommand{\taup}{\bar{\tau}'}
\newcommand{\Pme}{P_{me}}
\newcommand{\vtil}{\tilde{v}}

\newcommand{\todo}[1]{\textcolor{red}{[TODO: #1]}}
\newcommand{\rev}[1]{\textcolor{black}{#1}}
\newenvironment{revblock}{\color{black}}{}   

\begin{document}


\title{Shear--kSZ: A New Estimator for the Matter--Electron Power Spectrum from kSZ Tomography and Weak Lensing}

\author{Boryana Hadzhiyska}
\email{boryanah@ast.cam.ac.uk}
\affiliation{Institute of Astronomy, Madingley Road, Cambridge, CB3 0HA, UK}
\affiliation{Kavli Institute for Cosmology Cambridge, Madingley Road, Cambridge, CB3 0HA, UK}

\begin{abstract}
We propose a new estimator of the electron--matter power spectrum, $P_{me}(k)$, which
correlates the kinematic Sunyaev--Zel'dovich (kSZ) temperature with the product of the
line-of-sight velocity field and the weak-lensing convergence. Like the stacked kSZ
estimator, it factorizes into a calibratable velocity kernel times $P_{me}(k)$, but it
probes the full matter distribution rather than the gas around a biased tracer. It
therefore measures the baryonic suppression of the matter power spectrum, $S(k)$, one of
the dominant astrophysical systematics for Stage-IV cosmic shear. We derive an analytic
expression for the estimator, validate it against simulations at the per-cent level, and
show that its parity structure cancels cosmic microwave background (CMB) foregrounds.
\rev{We forecast the detectability with a realistic pipeline: velocities reconstructed from
the simulated tracers, with and without photometric-redshift errors; the kSZ signal of all
the ionized gas along the line of sight; a CMB map with Simons Observatory (SO) noise; and
LSST source samples free of intrinsic alignments. A DESI-like spectroscopic sample at
$0.5<z<1$ over the $20\%$ of the sky that DESI, LSST and SO share gives a $4$--$7\sigma$
detection. Photometric redshifts cost only $5$--$10\%$ of the significance, so LSST can
supply its own velocity tracers over its full overlap with SO: with tracers at $0.1<z<1.3$
and five source samples we find $10\sigma$ in an end-to-end simulation, and $13\sigma$ with
Advanced SO (16$\sigma$ if extending to $z < 1.5$). This corresponds to a
few-per-cent measurement of $S(k)$ at $k\simeq1$--$4\,h\,{\rm Mpc}^{-1}$, the scales that
matter most for cosmic shear.}
\end{abstract}

\maketitle

\section{Introduction}
\label{sec:intro}

Understanding the distribution of baryons in the universe is one of the central
challenges of modern observational cosmology
\citep[see, e.g.,][]{FukugitaPeebles2004, Shull2012, Nicastro2018, Macquart2020}.
While cosmological simulations and analytical models can track dark matter to
high accuracy, the behavior of the diffuse ionized gas that pervades the
cosmic web is far more uncertain, governed by complex feedback processes
associated with star formation and active galactic nuclei (AGN)
\citep{Chisari2019}.
These baryonic effects redistribute gas within and beyond the virial radii of
dark matter halos, suppressing the matter power spectrum on scales of
$k \sim 0.1$--$10\,h\,\mathrm{Mpc}^{-1}$ relative to a dark-matter-only
universe \citep{vanDaalen2011, Chisari2019}.
Quantifying this suppression accurately is critical for Stage-IV weak-lensing
surveys such as the Vera Rubin Observatory Legacy Survey of Space and Time
(LSST) \citep{Ivezic2019} and \textit{Euclid} \citep{Euclid2011}, whose statistical
precision will be limited by systematic uncertainties from baryonic physics
unless those uncertainties can be constrained observationally\rev{, or removed by
construction, e.g.\ by nulling the baryon-sensitive part of the shear signal with
cross-correlations against fast-radio-burst dispersion measures \citep{2025arXiv250919514L}}.

The kinematic Sunyaev--Zel'dovich (kSZ) effect \citep{1980ARA&A..18..537S} offers a
uniquely direct probe of the free-electron momentum field
\citep{2019SSRv..215...17M}.
Unlike the thermal SZ (tSZ) effect, the kSZ signal is spectrally degenerate
with the primary CMB and proportional to the optical depth times the
line-of-sight velocity, making it sensitive to the total ionized baryon
content regardless of gas temperature or metallicity.
Statistical kSZ estimators have been developed and applied to measure the
baryon profiles around galaxy halos, including the pairwise momentum estimator
\citep{2012PhRvL.109d1101H,2016MNRAS.461.3172S,2017JCAP...03..008D,2021PhRvD.104d3502C,2026PhRvD.113f3538G,2026PhRvD.113f3565H}, the
projected-field estimator \citep{2004ApJ...606...46D, 2016PhRvL.117e1301H,2021PhRvD.104d3518K},
velocity reconstruction and kSZ-tomography approaches
\citep{2009arXiv0903.2845H,Smith2018ksz,Munchmeyer2019,2025JCAP...05..057M,2025arXiv250621657H,2025PhRvL.134o1003L},
and stacked kSZ estimators around photometric and spectroscopic samples
\citep{2021PhRvD.103f3513S, Amodeo2021, 2025PhRvD.112h3509H, 2025PhRvD.112j3512R,2025PhRvD.111b3534H,2025PhRvD.112l3507H}.

A particularly fruitful class of estimators correlates a velocity template
derived from a spectroscopic galaxy survey with the kSZ temperature field.
The \textit{standard stacked kSZ estimator},
$\langle v_{\rm rec}\,\delta_g, T_{\rm kSZ}\rangle \propto
\langle v^2\rangle\,P^{ge}(k)$
\citep{2016PhRvD..93h2002S},
recovers the galaxy--electron cross-power spectrum $P^{ge}$ and has been
detected at high significance with current cutting-edge data
\citep{2026arXiv260419744Q,2026arXiv260419745H}.
However, this estimator probes $P^{ge}$ for a \emph{specific tracer} at a
specific mass scale; connecting it to the matter power spectrum suppression
requires modeling the galaxy--halo connection across a range of halo masses,
which introduces additional systematic uncertainty in deducing the suppression.

In this work we propose and validate a generalization of this estimator that
replaces the galaxy overdensity $\delta_g$ with the weak-lensing convergence
$\kappa$, which traces the projected matter density without any galaxy-bias
assumption.
Specifically, we consider the {\it shear--kSZ estimator},
\begin{equation}
\hat{C}_\ell^{TV_i} \equiv
\left\langle T\,,\, V_i \right\rangle_\ell,
\qquad
V_i(\nhat) = \frac{\vtil_i(\nhat)}{c}\,\kappa(\nhat),
\label{eq:estimator_def}
\end{equation}
where $T$ is the CMB temperature map containing the kSZ, $\vtil_i$ is a line-of-sight velocity template constructed from
spectroscopic tracers in a thin radial bin (with $i$ denoting which bin it is), 
and $\kappa$ is the lensing convergence from background galaxies. 
When the tracer bins are thin compared to the survey depth and the density
fields carry the small-scale power, the expectation value factorizes
(Section~\ref{sec:theory}) into
\begin{equation}
\langle \hat{C}_\ell^{TV_i}\rangle
= T_{\rm CMB}\;\mathcal{V}_i\;
W_\kappa(\chi_i)\,
\frac{P_{me}\!\left(k_\ell, z_i\right)}{\chi_i^2},
\label{eq:master_intro}
\end{equation}
where $k_\ell = (\ell+1/2)/\chi_i$, $W_\kappa$ is the lensing kernel, and
$\mathcal{V}_i$ is a velocity kernel, signifying the correlation between
reconstructed halo and true gas velocity, which we show can be
calibrated to high accuracy. This is highly analogous to the standard
stacked kSZ estimator, where
\begin{equation}
    \langle \hat{C}_\ell^{TV_g}\rangle
= T_{\rm CMB}\;\mathcal{V}_g\;
\frac{P_{ge}\!\left(k_\ell, z\right)}{\chi^2},
\label{eq:standard_stacked}
\end{equation} 
where $\mathcal{V}_g$ is the same halo--gas velocity kernel as above, but
averaged over the entire foreground sample.

As an internal consistency test, we cross-correlate the projected electron
optical-depth map directly with the projected matter overdensity in the
simulation; this cross $\tau$--matter reference agrees with the estimator,
confirming that $\hat{C}_\ell^{TV_i}$ captures $\Pme$.
The estimator thus provides direct observational access to the
matter--electron cross-power spectrum $\Pme(k)$, which encodes how the ionized
gas traces the total matter field. 
Crucially, we show in Section~\ref{sec:suppression} that an exact
decomposition of the matter field into cold dark matter and baryons relates
$\Pme$ to the baryonic suppression of the matter power spectrum,
$S(k) = P_{mm}/P_{mm}^{\rm DMO}$, up to a sub-percent dark-matter backreaction
term \citep{vanDaalen2020} and a small, bounded baryon auto-spectrum correction, eliminating the
halo-mass extrapolations required when the density proxy is a biased tracer.

\rev{We note that an independent study proposing a similar shear--kSZ
estimator appeared recently \citep{2026arXiv260729091G}. Their forecast, based on analytic signal templates and
covariances and on tomographic source bins combined with a nulling transformation,
is complementary to the simulation-based validation presented here, and the two
studies reach compatible conclusions; we comment on the points of contact in
Section~\ref{sec:discussion}.}

This paper makes three contributions beyond introducing the estimator.
First, we derive its expectation value from first principles, including the
line-of-sight velocity correlation structure, and identify an exact
sum rule: the line-of-sight velocity correlation function integrates to zero
over an infinite baseline, that dictates the tomographic design of the
measurement and the necessity of a calibrated velocity kernel.
Second, we validate the analytic model end-to-end against the simulation
measurement at the $\lesssim 5\%$ level over the signal-dominated scales,
using a velocity kernel measured from the same tracer catalog: the analogue
of the velocity-reconstruction transfer function familiar from stacked kSZ
analyses \citep{2021PhRvD.103f3513S}.
Third, we forecast the detectability with realistic noise: combining Simons
Observatory-like CMB data with Rubin/LSST lensing and a LRG
spectroscopic sample mimicking the Dark Energy Spectroscopic Instrument (DESI) \citep{2016arXiv161100036D}, jointly sharing $\fsky=0.2$ fraction of the sky, yields
\rev{$\mathrm{SNR}\approx 4$--$7$, depending on the fidelity of the velocity
reconstruction. When LSST supplies its own velocity tracers over its full
overlap with SO, with tracers at $0.1<z<1.3$ and several source samples, the
significance reaches $\approx10$, and $\approx13$ with Advanced SO. 
}

In this forecast, we use the AbacusSummit $N$-body simulation
\citep{2021MNRAS.508.4017M} and consider LRG-like reconstructed velocity field
in the redshift range $z=0.5$--$1$.
Luminous red galaxies (LRGs) are a natural choice because their spectroscopic
redshifts from DESI \citep{2023AJ....165...58Z} allow accurate line-of-sight
velocity reconstruction via the continuity equation, the same technique
employed in BAO reconstruction and stacked kSZ analyses
\citep{2015MNRAS.450.3822W,2024PhRvD.109j3533R,2024PhRvD.109j3534H,2026arXiv260723339O}.
We note that nothing in the estimator restricts the choice of tracer: the
Bright Galaxy Survey (BGS) at lower redshift, where background LSST source
galaxies are more abundant, is also viable (see
Section~\ref{sec:tracers} for the relevant trade-offs).

The paper is organized as follows.
Section~\ref{sec:theory} presents the theoretical framework: the estimator,
its expectation value, the velocity kernel and its calibration, the regime of
validity, and the connection between $\Pme$ and the matter power suppression.
Section~\ref{sec:methods} describes the simulations and the construction of
each field.
Section~\ref{sec:snr} details the covariance model and the Fisher forecast.
Section~\ref{sec:results} presents the validation of the analytic model, the
bias checks, and the SNR forecasts.
Section~\ref{sec:discussion} discusses the astrophysical implications and
concludes.

\section{Theoretical Framework}
\label{sec:theory}

There are three main ingredients in this estimator: a spectroscopic or photometric sample in the foreground, from which one can construct a line-of-sight velocity map, \rev{a galaxy weak-lensing (or CMB lensing) convergence map} in the background, the lensing kernel of which overlaps with the foreground sample, and a CMB temperature map, which includes a low-redshift kSZ contribution overlapping with the foreground sample. All also need to overlap on the sky.

\subsection{Fields}
\label{sec:theory_fields}

\textit{kSZ temperature.}
The kinematic Sunyaev--Zel'dovich temperature anisotropy along a line of sight
$\nhat$ is \citep{1980ARA&A..18..537S}
\begin{equation}
\frac{\Delta T_{\rm kSZ}(\nhat)}{T_{\rm CMB}}
= -\int_0^{\chi_*} d\chi\;
\taup(\chi)\,
\bigl[1+\delta_e(\chi\nhat)\bigr]\,
\frac{v_r(\chi\nhat)}{c},
\label{eq:ksz}
\end{equation}
where $v_r = \mathbf{v}\cdot\nhat$ is the line-of-sight peculiar velocity
(positive for recession), $\delta_e$ is the free-electron overdensity, and
\begin{equation}
\taup(\chi) \equiv \frac{d\bar\tau}{d\chi}
= \sigma_T\,\bar{n}_{e,0}\,(1+z)^2
\label{eq:taudot}
\end{equation}
is the differential mean optical depth per unit comoving distance, with
$\bar{n}_{e,0}$ the mean comoving free-electron number density and $\sigma_T$
the Thomson cross-section. 
We work in the optically thin regime, $e^{-\tau}\approx 1$.

\textit{Velocity template.}
For each radial bin $i$ of comoving width $\Delta\chi_{\rm bin}$, we construct
a two-dimensional line-of-sight velocity template $\vtil_i(\nhat)$ from the
tracer sample in that bin, smoothed with a Gaussian of comoving
radius $R_s$ (Section~\ref{sec:vfield}), to mimic the reconstructed velocity
field obtained from the standard reconstruction technique.
\rev{Our baseline template $\vtil_i$ is built from the smoothed true halo
velocities of the simulation; in a real analysis it would be the output of
continuity-equation velocity reconstruction applied to a spectroscopic (or
photometric) sample, and all statements below hold with $\vtil_i$ interpreted
as the reconstructed template. In Section~\ref{sec:recon} we also run the
reconstruction itself on the simulated halo catalog, with and without
photometric-redshift scatter, and use the reconstructed velocities directly, so
that the stochastic reconstruction noise enters both the response and the
covariance of the forecast.} 
The template normalization is not assumed: as shown below, the observable
depends only on calibrateable covariances of $\vtil_i$ with the true velocity
field, which play the role of the velocity-reconstruction transfer function in
stacked kSZ analyses.

\textit{Convergence.}
The lensing convergence for sources at comoving distance $\chi_s$ is
\begin{equation}
\kappa(\nhat) = \int_0^{\chi_s} d\chi\; W_\kappa(\chi;\chi_s)\,
\delta_m(\chi\nhat),
\label{eq:kappa_def}
\end{equation}
with the (flat-universe) kernel
\begin{equation}
W_\kappa(\chi;\chi_s) = \frac{3}{2}\,\Omega_m
\left(\frac{H_0}{c}\right)^{\!2} (1+z)\,
\chi\,\frac{\chi_s - \chi}{\chi_s}.
\label{eq:Wkappa}
\end{equation}
For an extended source distribution $dN/dz_s$, the kernel is replaced by its
source-averaged counterpart,
\begin{equation}
\bar{W}_\kappa(\chi) = \frac{\int dz_s\,\frac{dN}{dz_s}\,
W_\kappa\bigl(\chi;\chi(z_s)\bigr)}{\int dz_s\,\frac{dN}{dz_s}},
\label{eq:Wkappa_eff}
\end{equation}
which we use for the LSST-like configuration.
The velocity-weighted convergence field is then
$V_i(\nhat) = \bigl[\vtil_i(\nhat)/c\bigr]\,\kappa(\nhat)$, and the estimator
is the pseudo-$C_\ell$ cross-spectrum
\begin{equation}
\hat{C}_\ell^{TV_i} =
\frac{1}{2\ell+1}
\sum_{m=-\ell}^{\ell}
T^*_{\ell m}\,V_{i,\ell m},
\label{eq:estimator}
\end{equation}
with $T_{\ell m}$ denoting the harmonic space transform of the CMB temperature map.
\rev{A note on notation: throughout the paper, $\langle\cdots\rangle$ denotes an ensemble
average. The estimator itself, written $\langle X, Y\rangle_\ell$ with a comma, is the
pseudo-$C_\ell$ cross-spectrum of Eq.~\eqref{eq:estimator}, i.e.\ an average over $m$ at
fixed $\ell$ of a single realization, whose ensemble expectation is the cross-power spectrum.
The calibration matrix $A_{ij}$ of Eq.~\eqref{eq:Aij_def} is defined as a sky (pixel) average
over our full-sky, statistically isotropic maps; by ergodicity it estimates the same
ensemble covariance at zero angular lag, and we use the same bracket for both.}

\subsection{Expectation value}
\label{sec:theory_expectation}

The expectation of Eq.~\eqref{eq:estimator} is a four-point function of the
schematic form
$\langle (\delta_e\,v_r) , (\vtil_i\,\kappa)\rangle$.
Its leading Wick contractions are
\begin{equation}
\langle v_r\,\vtil_i\rangle\,\langle \delta_e\,\kappa\rangle
\;+\;
\langle \delta_e\,\vtil_i\rangle\,\langle v_r\,\kappa\rangle,
\label{eq:wick}
\end{equation}
plus connected non-Gaussian terms (see also Ref.~\citep{2026JCAP...01..015W} for derivation of 
the standard stacked kSZ estimator).
The two contractions have very different harmonic support.
Velocity fields are coherent on large scales: in the Limber regime the
projected velocity carries multipoles only up to
$L_v \sim \chi\,k_v \sim \mathcal{O}(10^2)$, where
$k_v \sim 0.1\,h\,{\rm Mpc}^{-1}$ bounds the support of the velocity power
spectrum (further reduced by the smoothing $R_s$).
Density fields carry power to arbitrarily small scales.
In the first contraction the two velocities pair at low multipole and the two
densities pair at the multipole of interest; in the second, both pairings are
density--velocity cross-correlations, which die off above
$\ell \sim$ a few hundred.
For $\ell \gg L_v$ the first term therefore dominates, and the product
structure of $T$ and $V_i$ implies a convolution of the form,
\begin{equation}
\langle\hat{C}_\ell^{TV_i}\rangle
= \int\!\frac{d^2 L}{(2\pi)^2}\,
C^{v\vtil_i}_{L}\;
C^{\,\taup\delta_e\times\kappa}_{|\boldsymbol{\ell}-\mathbf{L}|},
\label{eq:convolution}
\end{equation}
written here in the flat-sky approximation.
Since the density cross-spectrum is smooth over the width $L_v$ of the
velocity kernel, similarly to the standard stacked kSZ estimator,
the convolution collapses for $\ell \gg L_v$ to the velocity
covariance at zero angular lag multiplying the density cross-spectrum.
Restoring the line-of-sight structure and applying the Limber approximation to
the $\langle\delta_e\,\kappa\rangle$ pairing (which collapses the lensing
integral onto the kSZ shell), we obtain the master formula
\begin{align}
\langle\hat{C}_\ell^{TV_i}\rangle
&= \rev{-}\frac{T_{\rm CMB}}{c^2}
\int d\chi\;
\taup(\chi)\,
\xi^v_i(\chi)\;
W_\kappa(\chi)\,
\frac{\Pme\!\left(k_\ell; z(\chi)\right)}{\chi^2},
\label{eq:master_exact}
\end{align}
\rev{(the overall minus sign is that of Eq.~\ref{eq:ksz}; it is absorbed into the
velocity kernel $\mathcal{V}_i$ defined in Eq.~\ref{eq:Vi_def} below, so that
Eqs.~\ref{eq:master_intro}, \ref{eq:master_factorized} and \ref{eq:master_discrete}
are mutually consistent)}
with $k_\ell = (\ell+1/2)/\chi$ and
\begin{equation}
\xi^v_i(\chi) \equiv
\bigl\langle v_r(\chi\nhat)\;\vtil_i(\nhat)\bigr\rangle
\label{eq:xiv_def}
\end{equation}
the covariance between the true line-of-sight velocity at distance $\chi$ and
the template of bin $i$, evaluated along the same line of sight.
Because $\xi^v_i(\chi)$ is sharply peaked around the bin (with structure
and width we discuss and quantify next) 
while $W_\kappa$, $\Pme$, and $\chi^{-2}$ vary slowly over its
support, Eq.~\eqref{eq:master_exact} roughly factorizes to the form quoted in
Eq.~\eqref{eq:master_intro},
\begin{eqnarray}
\langle\hat{C}_\ell^{TV_i}\rangle
\simeq T_{\rm CMB}\;\mathcal{V}_i\;
W_\kappa(\chi_i)\,\Cme(\chi_i), \nonumber
\\
\Cme(\chi) \equiv \frac{\Pme(k_\ell; z)}{\chi^2},
\label{eq:master_factorized}
\end{eqnarray}
where the velocity kernel is
\begin{equation}
\mathcal{V}_i \equiv - \frac{1}{c^2}\int d\chi\;\taup(\chi)\,\xi^v_i(\chi).
\label{eq:Vi_def}
\end{equation}
All of the estimator's dependence on the velocity field, the template
construction (smoothing, weighting, reconstruction fidelity), and the survey
radial geometry is contained in the single number $\mathcal{V}_i$ per bin;
the scale dependence is carried entirely by $\Pme(k_\ell)$.
This separation is the central property of the estimator: $\mathcal{V}_i$ can
be calibrated e.g., via linear theory or simulations
(Section~\ref{sec:theory_calibration}), while $\Pme$ is the
cosmological target. We note that this is very similar to the analytical 
expression for the standard stacked estimator derived in Refs.~\citep{2025arXiv251214625H,2026arXiv260419744Q,2026arXiv260419745H},
swapping $P_{ge} \rightarrow W_\kappa P_{me}$.

\subsection{The line-of-sight velocity kernel}
\label{sec:theory_velkernel}

The structure of $\xi^v_i(\chi)$ is set by the line-of-sight velocity
correlation function and can be estimated analytically or via simulations.
For a template that averages the velocity over bin $i$,
\begin{equation}
\xi^v_i(\chi) = \mathcal{A}_i\,
\frac{1}{\Delta\chi_{\rm bin}}
\int_{{\rm bin}\,i} d\chi'\;
\Psi^{(R_s)}_\parallel(\chi - \chi'),
\label{eq:xiv_structure}
\end{equation}
where $\mathcal{A}_i$ is a normalization absorbing the template construction
(Section~\ref{sec:theory_calibration}) and $\Psi^{(R_s)}_\parallel(r)$ is the
line-of-sight--line-of-sight velocity correlation function at radial
separation $r$, smoothed on scale $R_s$. In linear theory,
\begin{eqnarray}
\Psi^{(R_s)}_\parallel(r) =
\frac{(f a H)^2}{4\pi^2}
\!\int\! dk\,P(k) \times \nonumber
\\
\!\int_{-1}^{1}\! d\mu\;\mu^2
\cos(k\mu r)\,
e^{-k^2(1-\mu^2)R_s^2/2},
\label{eq:Psi_par}
\end{eqnarray}
which reduces at $R_s = 0$ to the standard result
$\Psi_\parallel(r) = (faH)^2/(2\pi^2)\int dk\,P(k)\,
[\,j_0(kr) - 2j_1(kr)/(kr)\,]$, with
$\Psi_\parallel(0) = \sigma_{v,\rm 1D}^2$ \citep{Gorski1988}.

Two properties of $\Psi_\parallel$ dictate the design of the measurement.
First, its correlation length is $\sim 40\;\mpc$: this motivates the choice of
bin width, since much wider bins average the velocity toward zero while much
narrower bins add strongly covariant tomographic slices with gaining very little
independent information.
Second, and more consequentially, the line-of-sight velocity correlation obeys
an exact sum rule,
\begin{equation}
\int_{-\infty}^{\infty} dr\;\Psi_\parallel(r) = 0,
\label{eq:sumrule}
\end{equation}
which follows immediately in Fourier space: the full-line integral projects
onto $k_\parallel = 0$ modes, which carry no line-of-sight velocity.
The correlation function therefore has a positive core of width
$\sim 40\;\mpc$ and a shallow negative tail extending to several hundred
$\mpc$.
Three consequences follow:
(i) a single unbinned velocity weight over a thick survey volume yields
\rev{almost} no signal\rev{. With tomographic binning, the response of shell $i$
is non-zero only because the line-of-sight weight
$\taup W_\kappa \Pme/\chi^2$ in Eq.~\eqref{eq:master_exact} varies across the
radial support of $\xi^v_i$, so that the positive core and the negative tail
do not cancel exactly. The binning therefore determines the velocity kernel
but does \emph{not} by itself localize the kSZ contribution to the tracer
shell: because $\kappa$ is a broad projection, ionized gas along the entire
line of sight contributes to $\mathcal{V}_i$ through the extended tail of the
velocity correlation. We quantify this contribution, which reduces the signal
of the interior shells by $\sim30\%$, in Appendix~\ref{sec:wings}};
(ii) $\mathcal{V}_i$ is intrinsically sensitive to the detailed shape of the
velocity correlation, including its mildly non-linear tail, and should be
calibrated or predicted analytically; and
(iii) the radial edges of the tracer volume need special attention as part of their
negative tail \rev{and, for the outermost shells, part of their positive core}
fall outside the tracer volume and thus require additional calibration. 
For ease, we drop them in this analysis (see Appendix~\ref{app:edges}).
We verify all of these effects in Section~\ref{sec:validation}.

\subsection{Calibration of the velocity kernel}
\label{sec:theory_calibration}

In practice, we can model the velocity correlation function 
$\xi^v_i(\chi)$ analytically or via simulations. In this work, we compare both.
Because the template enters the estimator linearly and only through its
low-multipole modes, the discrete counterpart of
Eqs.~\eqref{eq:master_exact}--\eqref{eq:Vi_def} is
\begin{equation}
\langle\hat{C}_\ell^{TV_i}\rangle
\approx \rev{-}\frac{T_{\rm CMB}}{c^2}
\sum_j \Delta\tau_j\;A_{ij}\;
W_\kappa(\chi_j)\,\Cme(\chi_j),
\label{eq:master_discrete}
\end{equation}
where the sum runs over all radial bins $j$ covered by the kSZ integration
range \rev{(i.e.\ in principle the full line of sight, including bins outside the
tracer volume, for which $A_{ij}$ is not directly measurable from the tracers
and must be supplied by linear theory or by a reconstruction over a wider
volume; see Appendix~\ref{sec:wings})}, $\Delta\tau_j$ is the mean optical depth through bin $j$, and
\begin{equation}
A_{ij} \equiv
\bigl\langle \vtil_i(\nhat)\,\bar{v}_j(\nhat) \bigr\rangle
\label{eq:Aij_def}
\end{equation}
is the sky-averaged covariance between the template of bin $i$ and 
the line-of-sight (projected and unsmoothed) velocity of bin $j$. We note that adopting the projected field is a slight simplification of the analytical formula, but it yields sufficient accuracy for the purposes of this study.
In a real analysis, the band matrix $A_{ij}$ can either be calculated
analytically or calibrated via simulations, and is analogous to the
normalization constant $r$ in stacked kSZ analyses. In simulations, 
one can compute it directly using the velocity field.
Equation~\eqref{eq:master_discrete} has two important practical features.
First, the effects of reconstruction noise,
smoothing suppression, or the sampling of the velocity field by the discrete
tracers are all absorbed into $A_{ij}$.
Second, because $A_{ij}$ is measured from the same realization, it captures the
sample variance of the long-wavelength radial velocity modes, which modulate
the amplitude of the estimator coherently across neighboring bins and across
all multipoles (Section~\ref{sec:validation}).
The diagonal $A_{ii}$ encodes the effective velocity variance of the template
(including reconstruction noise and smoothing losses), while the off-diagonal
encodes the correlation structure of Eq.~\eqref{eq:xiv_structure}. We note 
that the correlation structure can also be studied directly using the 
reconstructed velocity field alone with some additional modeling 
\citep[see also Ref.~][for similar derivations]{2025arXiv251115701M}.

\subsection{Regime of validity}
\label{sec:theory_validity}

Equation~\eqref{eq:master_discrete} rests on two approximations, both of which
fail only at low multipoles.
First, the collapse of the convolution in Eq.~\eqref{eq:convolution} requires
$\ell \gg L_v$; below a few hundred, the measured spectrum flattens toward a
floor set by the convolution rather than tracking $\Cme$.
Second, the neglected contraction
$\langle\delta_e\,\vtil_i\rangle\langle v_r\,\kappa\rangle$ in
Eq.~\eqref{eq:wick} consists of two density--velocity cross-spectra and
contributes only at $\ell \lesssim$ a few hundred, where it adds positive
power.
Both effects push the measurement above the factorized prediction at low
$\ell$, consistent with what we observe in
Section~\ref{sec:validation}.
We therefore restrict quantitative use of the model to
$\ell \gtrsim 500$, 
noting that the forecast SNR is in any
case dominated by $\ell \sim 10^3$--$5\times10^3$.

\subsection{From $\Pme$ to the matter power suppression}
\label{sec:suppression}

The quantity of greatest interest for weak-lensing cosmology is the baryonic
suppression of the matter power spectrum,
$S(k) \equiv P_{mm}(k)/P_{mm}^{\rm DMO}(k)$.
We now show that $\Pme$ determines $S(k)$ up to small corrections. 
This is also verified on a hydrodynamical simulation in App.~\ref{app:suppression}.

Decompose the matter field into cold dark matter and baryons,
$\delta_m = f_c\,\delta_c + f_b\,\delta_b$, with
$f_b = \Omega_b/\Omega_m \approx 0.157$ and $f_c = 1 - f_b$.
Then, exactly,
\begin{equation}
P_{mm} = f_c^2\,P_{cc} + 2 f_b\,P_{mb} - f_b^2\,P_{bb}.
\label{eq:decomposition}
\end{equation}
For a fully ionized medium, and neglecting for the moment the small fraction
of baryons locked in stars (we could equivalently set it to a known
constant and reach the same conclusion), the free electrons trace the baryons,
$\delta_b \simeq \delta_e$, so $P_{mb} \simeq \Pme$: the second term is
precisely what shear--kSZ measures.
The suppression is therefore
\begin{equation}
S(k) = f_c^2\,\frac{P_{cc}}{P_{mm}^{\rm DMO}}
\;+\;
\frac{2 f_b\,\Pme - f_b^2\,P_{ee}}{P_{mm}^{\rm DMO}}.
\label{eq:suppression_exact}
\end{equation}
The first term is controlled by the dark-matter backreaction: hydrodynamical
simulations consistently find
$P_{cc}/P_{cc}^{\rm DMO} = 1 + \mathcal{O}(0.1\%)$ on the scales of interest
\citep{vanDaalen2020, Chisari2019},
so $f_c^2 P_{cc}/P_{mm}^{\rm DMO} \simeq f_c^2$ to sub-percent accuracy in
$S$ (see App.~\ref{app:suppression} and Ref.~\citep{2025A&A...697A..63O}).
The second term gives the main baryon-correction to the matter suppression.
The last term carries the small weight $f_b^2 \approx 0.025$ and is bounded by
the Cauchy--Schwarz inequality, $P_{ee} \ge \Pme^2/P_{mm}$, with near-equality
until $k \lesssim 10 \ h/{\rm Mpc}$ \citep{2025MNRAS.538.1415S} because the electron--matter 
correlation coefficient is close to unity on
these scales. Substituting the bound closes the system in terms of measured
quantities alone. Thus, one can obtain a sub-percent prediction of $S(k)$
if $P_{me}$ is known. In fact as baryons make up only 15--16\% of the total
matter, a 5\% measurement of $P_{me}$ \rev{at a given $k$} translates into a $\lesssim 1\%$ measurement
of $S(k)$ \rev{at that $k$}. In fact, since what matters is the deviation from $S(k) = 1$, in the presence of strong feedback the 
significance is even larger, as shown in Table~\ref{tab:propagation} and App.~\ref{app:suppression_propagation}.

Corrections from the stellar baryon fraction $f_\star$ (electrons trace only
the ionized gas, $\delta_e = \delta_{\rm gas}$ up to helium and ionization
factors absorbed in $\bar{n}_{e,0}$) enter at
$\mathcal{O}(f_\star f_b) \sim 1\%$ of $S$ and can be modelled with standard
stellar-mass functions to sub-percent accuracy.

As a proof of concept, we demonstrate in Appendix~\ref{app:suppression} that
this inference recovers the true matter power suppression of the MillenniumTNG
simulation from $P_{me}$ alone to better than $0.5\%$, with the biggest
effect coming from ignoring the dark-matter
backreaction (Fig.~\ref{fig:backreaction}), which if accounted for improves the accuracy to 0.2\%.

We emphasize the contrast with the stacked kSZ estimator,
$\langle v_{\rm rec}\,\delta_g, T_{\rm kSZ}\rangle \propto
\langle v^2\rangle P^{ge}$:
there the density proxy is a biased tracer, and inferring $S(k)$ requires
modelling the galaxy--halo connection and extrapolating across halo mass.
The shear--kSZ estimator replaces $\delta_g$ with $\kappa$, so the density leg is the total matter
field and Eq.~\eqref{eq:suppression_exact} applies directly.

\subsection{Robustness to foregrounds and lensing systematics}
\label{sec:robustness}

Three properties and choices of our estimator make it robust to
foregrounds and systematics, allowing us to measure the ionized-gas--matter
cross-power directly.

\emph{Foregrounds cancel by velocity parity.}
The expectation value of $\hat{C}_\ell^{TV_i}$ is a bispectrum of the form
$\langle X\,\vtil_i\,\kappa\rangle$, where $X$ is whatever sources the
temperature map.
Under a reversal of the peculiar-velocity field, $\mathbf{v} \to -\mathbf{v}$,
the template $\vtil_i$ is odd, while every velocity-independent sky component:
primary CMB, tSZ, cosmic infrared background, radio point sources, Galactic
dust, and the convergence $\kappa$ are even.
The ensemble average of an odd quantity vanishes:
$\langle X\,\vtil_i\,\kappa\rangle = 0$ for any $X$ that does not know about
the line-of-sight velocity, regardless of how strongly $X$ correlates with the
density field (the tSZ and CIB correlate with $\kappa$ very strongly, and
still drop out).
Components linear in $v_r$ survive, of which the kSZ is by far the
dominant one.
Foregrounds therefore add variance but no bias, exactly as in stacked kSZ
analyses, which is the defining advantage of velocity-weighted estimators.

\emph{No intrinsic alignments or boost factors.}
IA contamination and boost-factor corrections in galaxy--galaxy lensing arise
from physical association between sources and lenses, i.e.\ from overlap of
the source redshift distribution with the lens volume.
Our high-$z$ source selection (Section~\ref{sec:kappa_lsst}) has no support at
$z \le 1$ by construction, removing both effects entirely rather than modelling them.
This becomes progressively easier at lower lens redshift, where a large fraction of the
source population lies behind the tracer volume. This is a further argument for a
BGS-based application of the estimator (Section~\ref{sec:tracers}).

\emph{No galaxy bias on the density leg.}
The matter proxy is $\kappa$; no bias expansion, miscentering, halo occupation
distribution, or halo-mass
extrapolation enters the interpretation of the density side of the
measurement, in contrast to $P^{ge}$-based estimators.
The only astrophysical calibration is the velocity kernel $A_{ij}$, which is a
large-scale, linear-regime quantity.

\section{Simulations and Field Construction}
\label{sec:methods}

\subsection{AbacusSummit simulation}
\label{sec:abacus}

All fields are constructed from the AbacusSummit suite \citep{2021MNRAS.508.4017M},
specifically the \texttt{AbacusSummit\_huge\_c000\_ph201} box: a
$(7.5\,h^{-1}\mathrm{Gpc})^3$ volume with $8640^3$ particles,
with particle mass $\approx 5.6\times10^{10}\,\msol$,
at the Planck 2018 cosmology \citep{Planck2018params}
($\Omega_m = 0.3152$, $h = 0.6736$).
The associated full-sky particle light cone provides the matter field in thin
spherical shells (one per simulation time step, of width
$\approx 6\;\mpc$ at the redshifts of interest), the halo light-cone
catalogs \citep{Hadzhiyska2022lc}
provide LRG-like tracers with positions and peculiar velocities over
$z=0.5$--$1$, and the AbacusLensing suite \citep{AbacusLensing} of convergence planes
from the same realization provides the $\kappa$ maps. We note that all catalogs are full-sky until $z \approx 2.35$.
All maps are produced in the \HEALPIX\ \citep{Gorski2005} ring scheme at $N_{\rm side}=4096$
(resolution $\approx 0.86'$), with $\lmax=8000$.

\subsection{kSZ temperature map}
\label{sec:ksz_map}

The kSZ map is built shell by shell from the particle light cone.
For each light-cone step $s$ we form the projected matter overdensity
$\delta_{m,s}(\nhat)$ from the particle counts (note that this is the dark-matter-only matter field), convert it to an electron
overdensity in harmonic space using a gas transfer function calibrated on the
MillenniumTNG hydrodynamical simulation
\citep{MTNGHernandezAguayo2023, MTNGPakmor2023,2023MNRAS.526..369H,2025MNRAS.538.1415S,2026PhRvD.113f3558L},
\begin{equation}
\delta_{e,s}(\vk) = T_{\rm gas}(k)\,\delta_{m,s}(\vk),
\qquad
T_{\rm gas}(k) = \sqrt{\frac{P_{\rm gas}(k)}{P_{\rm dm}(k)}}\bigg|_{\rm MTNG},
\label{eq:transfer}
\end{equation}
evaluated at the Limber wavenumber $k = (\ell+1/2)/\chi_s$, and weight by the
shell's mean optical depth $\Delta\tau_s$ (Eq.~\ref{eq:taudot}, integrated
across the shell assuming fully ionized hydrogen and helium) and the
line-of-sight peculiar velocity:
\begin{equation}
T_{\rm kSZ}(\nhat) = - T_{\rm CMB}
\sum_s \Delta\tau_s
\bigl[1 + \delta_{e,s}(\nhat)\bigr]\,
\frac{v_{r,s}(\nhat)}{c}.
\label{eq:ksz_map}
\end{equation}
\rev{For convenience, all simulated kSZ maps used in this paper are generated with the
opposite overall sign, $T_{\rm kSZ}\to -T_{\rm kSZ}$, so that the measured spectra
$\hat{C}_\ell^{TV_i}$ shown in the figures are positive; equivalently, the figures show
$-\hat{C}_\ell^{TV_i}$ in the convention of Eqs.~\eqref{eq:ksz} and \eqref{eq:master_exact}.
This overall sign is immaterial for the estimator and for all signal-to-noise statements.
Note also that the sum in Eq.~\eqref{eq:ksz_map} runs over the light-cone shells
covering the tracer volume $z=0.5$--$1$ only (see Appendix~\ref{sec:wings} for
the contribution of the electrons outside this range).}
We note that the choice of gas-painting technique -- whether it be
a transfer function, a baryon correction
model, or a direct hydrodynamical simulation, 
does not affect the validity of our estimator or our conclusions.

\subsection{Lensing convergence maps}
\label{sec:kappa}

We consider two configurations for the source convergence map $\kappa$.
A remark on observables: lensing surveys measure the shear field
$\gamma$, not $\kappa$ directly; on the small scales and (near-)full-sky
geometry relevant here the two carry identical information, related by the
standard Kaiser--Squires spin lowering, $\kappa_{\ell m} =
\gamma^{E}_{\ell m}$ with negligible for our application large-scale 
($\ell\sim10$) corrections, and convergence
maps are routinely reconstructed from shear catalogs.
We therefore phrase the estimator in terms of $\kappa$ throughout, and stress
that our estimator, or its significance of detection is unchanged if we
swap $\kappa \rightarrow \gamma^{E}$.

\subsubsection{Single source plane at $z_s = 1.9$}
\label{sec:kappa_z19}

In the simplest case we use the AbacusLensing convergence map at a single
source redshift $z_s = 1.9$, providing a high signal-to-noise reference with
no shape noise.

\subsubsection{LSST-like high-redshift source sample}
\label{sec:kappa_lsst}

For a realistic forecast we construct an effective $\kappa$ map for a
high-redshift LSST source selection, designed so that the source distribution
has negligible support within the lens range $z=0.5$--$1$: with disjoint lens and source distributions,
intrinsic alignments and boost-factor corrections are absent by construction
(Section~\ref{sec:robustness}), at the cost of a reduced source density.
Starting from the full LSST Y10 parent distribution, modelled as a Smail-type
law $dN/dz \propto z^2 \exp[-(z/z_0)^{\alpha}]$ with $(z_0, \alpha) = (0.11,
0.68)$ \citep{DESC_SRD}, we select a photometric bin $z_{\rm ph} \in [1.5,
2.6]$ with Gaussian photo-$z$ scatter $\sigma_z = 0.05(1+z)$, yielding the
true-redshift distribution
\begin{equation}
p(z) \propto \frac{dN}{dz}\,
\frac{1}{2}\left[
\mathrm{erf}\!\left(\frac{z - z_{\rm ph}^{\rm lo}}{\sqrt{2}\,\sigma_z}\right)
- \mathrm{erf}\!\left(\frac{z - z_{\rm ph}^{\rm hi}}{\sqrt{2}\,\sigma_z}\right)
\right],
\label{eq:dndz}
\end{equation}
a realistic asymmetric shape comparable to the highest tomographic source bins
of LSST analyses \citep[e.g.,][Fig.~1]{Nicola2023}, with mean redshift
$\langle z\rangle \approx 1.89$ and negligible ($<10^{-3}$) weight at
$z \le 1$. The effective convergence is the plane-weighted sum
$\kappa_{\rm LSST}(\nhat) = \sum_i w_i\,\kappa_{z_i}(\nhat)$ over the
AbacusLensing planes ($z = 0.15$--$2.35$), with $w_i$ the integral of $p(z)$
across each plane's slice and the $z>2.35$ tail folded into the last plane
(a $<1\%$ effect); the corresponding theory kernel is the source-averaged
$\bar{W}_\kappa$ of Eq.~\eqref{eq:Wkappa_eff} with the same weights.

To this signal we add Gaussian shape noise with a white angular power spectrum
\begin{equation}
N_\ell^{\kappa} = \frac{\sigma_e^2}{\bar{n}_{\rm sr}},
\label{eq:kappa_noise}
\end{equation}
where $\sigma_e = 0.26$ \rev{is the per-component ellipticity dispersion
\citep{DESC_SRD},} and $\bar{n}$ is the effective number density of the
\emph{selected subset}: the photometric window retains a fraction
$\approx 0.17$ of the
$27\,\mathrm{arcmin}^{-2}$ Y10 parent sample, i.e.\
$n_{\rm eff} \approx 4.6\,\mathrm{arcmin}^{-2}$ \citep[values taken from Ref.~][]{DESC_SRD}.
Thus, we make sure that the kernel and the noise describe the same galaxies: 
the high lensing efficiency (relative to the LRG sample)
of the $z\sim1.75$ selection comes at the price of $\sim5\times$ larger
shape-noise power than the full sample. 

We note that our choice of a background source sample is somewhat arbitrary; 
likely, a more optimal sample yielding higher signal-to-noise can be constructed, 
but we leave this for future work. Furthermore, while in our forecast we adopt the 
Y10 number density of sources, \rev{earlier data releases lose less than one
might expect: the Y1 (Y3) shape noise of the same selection lowers the
significance by $\approx35\%$ ($\approx15\%$), because the CMB temperature map
dominates the noise (Section~\ref{sec:snr_results}).}
Lower-redshift source samples would
present lower shape-noise and yield a higher signal-to-noise when paired with
a lower-redshift reconstructed velocity field.

\subsection{Observed CMB temperature map}
\label{sec:T_obs}

In the two ``observed CMB'' scenarios we replace the ideal kSZ-only temperature
field with a realistic observed map,
\begin{equation}
T_{\rm obs}(\nhat) = B(\nhat) * \bigl(T_{\rm CMB}(\nhat) + T_{\rm kSZ}(\nhat)\bigr)
+ n(\nhat),
\label{eq:T_obs}
\end{equation}
where $B_\ell = e^{-\ell(\ell+1)\sigma_b^2/2}$ is the harmonic space
transform of the beam with
$\sigma_b$ corresponding to a FWHM of $1.6'$ (representative of the Simons
Observatory (SO) Large Aperture Telescope (LAT) \citep{2019JCAP...02..056A}),
$T_{\rm CMB}$ is a Gaussian realization of the lensed primary CMB with the
\textit{Planck} 2018 power spectrum, and $n$ is a Gaussian noise realization
drawn from the Simons Observatory LAT component-separated (ILC, Deproj-0)
noise curve \citep{2019JCAP...02..056A}, rescaled by a factor 
$1/1.5$ to reflect the enhanced SO LAT noise \citep{2025JCAP...08..034A}, 
corresponding to roughly 8 $\mu K$-arcmin white-noise floor.
\rev{The public ILC noise curves are beam-deconvolved; since the signal in
Eq.~\eqref{eq:T_obs} carries the beam, the noise power at the map level is
$B_\ell^2 N_\ell$, and this is the convention we adopt in the covariance
(Section~\ref{sec:covariance}).}
The beam suppresses both the primary CMB and the kSZ signal, so the expected
estimator value for the observed case is
\begin{equation}
\langle\hat{C}_\ell^{TV_i}\rangle_{\rm obs}
= B_\ell\,\langle\hat{C}_\ell^{TV_i}\rangle_{\rm raw},
\label{eq:beam_correction}
\end{equation}
which we verify in Section~\ref{sec:bias_check}.
When comparing measurements to the analytic model we additionally account for
the \HEALPIX\ pixel window at $N_{\rm side}=4096$, which enters squared (once
through each map).

\subsection{Velocity templates}
\label{sec:vfield}

The halo velocity field is obtained directly from the AbacusSummit halo
light-cone catalogs, which provide 3D peculiar velocities for each halo.
We project these onto the line of sight and assign halos to 24 radial bins of
uniform comoving width $\Delta\chi_{\rm bin} = 40\,\mpc$ spanning
$z = 0.5$--$1$, roughly matching the correlation length of the line-of-sight velocity
field (Section~\ref{sec:theory_velkernel}).

Within each bin, the mean line-of-sight velocity of the halos in each
\HEALPIX\ pixel is deposited on the sky and the resulting map is smoothed with
a Gaussian kernel of comoving radius $R_s = 15\,\mpc$ (an angular scale
$R_s/\chi_i$ that varies with the distance to the bin centre) to mimic the
effective of smoothing applied in standard velocity reconstruction (e.g.,
\texttt{pyrecon}).
The smoothing suppresses shot noise from the discrete halo positions and
mimics the resolution of continuity-equation velocity reconstruction.
Pixels containing no halos are set to zero prior to smoothing; the resulting
dilution of the template amplitude by the occupied-pixel fraction
($f_{\rm occ}\sim 1$--$2\%$ per bin at $N_{\rm side}=4096$), together with the
smoothing suppression (measured to be $0.51$; the transverse Gaussian alone
predicts $0.70$ in linear theory, the difference arising from harmonic
band-limiting, pixel sampling, and small-scale velocity dispersion),
is absorbed by the calibration matrix $A_{ij}$ of
Eq.~\eqref{eq:Aij_def}, which we measure from the same catalog by
cross-correlating the smoothed template of bin $i$ with the deposited
(unsmoothed) velocities of every bin $j$. 
In principle, this can be evaluated even more accurately by using the full
particle velocity field, but for our purposes of matching the measurement 
at the 5\% level, this is sufficient. %

Our smoothed-halo-velocity construction is a stand-in for continuity-equation
velocity reconstruction, in which the 3D tracer overdensity is smoothed
(typically on $10$--$15\;\mpc$), divided by the linear bias and the
appropriate growth factors, and converted to a line-of-sight velocity: the
same machinery used in BAO reconstruction and stacked kSZ analyses.
\rev{Standard reconstruction achieves one-point correlation coefficients
$r = \langle v_{\rm rec} v_h\rangle / (\sigma_{v_{\rm rec}}\sigma_{v_h})
\approx 0.6$--$0.7$ with the true halo velocities for DESI-like samples
(we measure $r=0.65$ for the reconstruction of Section~\ref{sec:recon}); the
smoothed-true-velocity proxy has, by construction, no stochastic noise, and
the comparison of the two at the level of the response and of the template
auto-spectrum is deferred to Section~\ref{sec:recon}.}
We note that using non-linear reconstruction could potentially increase 
the signal-to-noise by 20--30\%.

\rev{The smoothed-true-velocity proxy reproduces the mean response of a
reconstructed template but not necessarily its noise: the stochastic part of
the reconstruction, uncorrelated with the true velocities, does not enter
$A_{ij}$ but does enter the template auto-spectrum $C_\ell^{V_iV_j}$ in the
covariance. We therefore also run continuity-equation reconstruction
(\texttt{pyrecon}, multigrid solver, smoothing radius $12.5\;\mpc$, local line
of sight, $f$ and $b$ fixed at the sample's effective redshift) on the halo
light-cone catalog with observed (redshift-space) positions, with and without
a photometric-redshift scatter $\sigma_z=0.02(1+z)$, and build a second set of
templates from the reconstructed line-of-sight velocities
$v_{\rm rec}=aHf\,\Psi_\parallel$, binning halos by their observed distance.
The comparison of the two sets of templates is presented in
Section~\ref{sec:recon}.}
For a data analysis, $A_{ij}$
would be measured from the reconstruction pipeline applied to mocks and data. 
The estimator is insensitive to the overall units of the template,
as they are absorbed by $A_{ij}$. Nonetheless, in the standard data analysis
we convert the displacements returned by the reconstruction pipeline  
$\psi = v\,(1+z)/H(z)$ into velocities, treating the pipeline and
conversion self-consistently (i.e., assuming the same cosmology).

\begin{figure}[t]
\centering
\includegraphics[width=\linewidth]{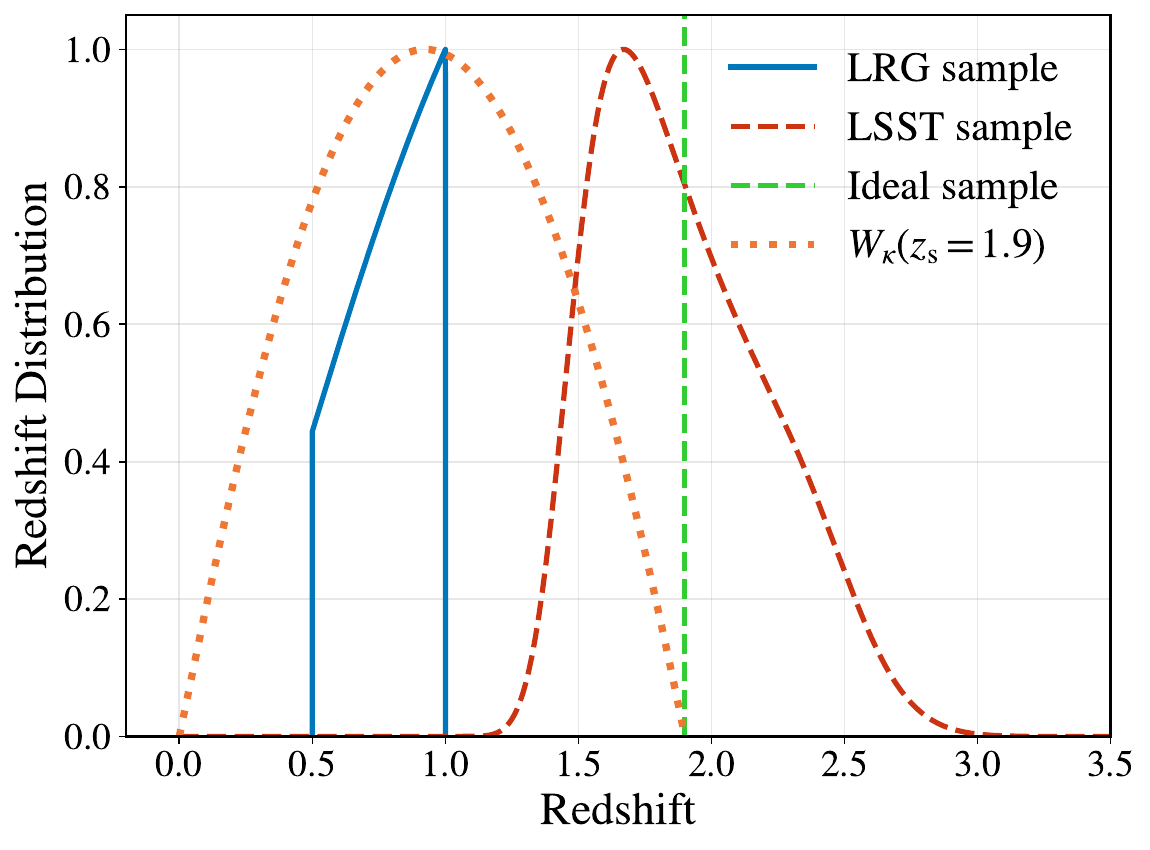}
\caption{Redshift distributions of the ingredients entering the estimator.
The velocity field is reconstructed from an LRG-like spectroscopic sample
spanning $0.5 < z < 1$ (solid blue), which also hosts the electrons sourcing
the kSZ signal. The matter leg is provided by the lensing convergence: we
contrast an idealized single source plane at $z_{\rm s} = 1.9$ (vertical
dashed green) with a realistic LSST-like source distribution (dashed red),
whose median redshift lies close to the ideal plane. The lensing efficiency
kernel $W_\kappa(z)$ for $z_{\rm s} = 1.9$ (dotted orange) peaks near
$z \simeq 0.9$, providing near-maximal overlap with the LRG footprint in
redshift. All curves are normalized to unit maximum.}
\label{fig:nz}
\end{figure}

Figure~\ref{fig:nz} summarizes the redshift configuration of the measurement.
The estimator requires three overlapping ingredients: the reconstructed
velocity field, the free electrons producing the kSZ temperature
fluctuations, and the matter fluctuations probed by CMB or galaxy lensing.
The first two are supplied by the same LRG-like sample at $0.5 < z < 1$,
so the cross-correlation is localized to this shell by construction. The
matter leg, by contrast, is an integrated quantity: the convergence for
sources at $z_{\rm s}$ weighs the intervening matter by the efficiency
kernel $W_\kappa(z) \propto \chi(z)\,[\chi(z_{\rm s}) - \chi(z)] /
\chi(z_{\rm s})\,(1+z)$, which is broad and peaks roughly midway in
comoving distance to the sources. For $z_{\rm s} = 1.9$ this peak falls
at $z \simeq 0.9$, squarely within the LRG shell, making such sources
nearly optimal for isolating $P_{m e}$ at the electron redshifts. An
LSST-like source sample, with $\mathrm{d}n/\mathrm{d}z$ peaking at
$z \simeq 1.75$ and a tail extending beyond $z \sim 3$, closely approximates
this ideal configuration: since $W_\kappa$ varies slowly with $z_{\rm s}$
for sources well behind the lens shell, the effective kernel obtained by
averaging over the LSST distribution differs only modestly from the
single-plane case, and essentially all sources lie behind the LRG volume,
minimizing source--lens overlap and the associated intrinsic-alignment and
boost-factor complications.

\section{Signal-to-Noise Estimation}
\label{sec:snr}

\subsection{Gaussian covariance}
\label{sec:covariance}

The covariance between $\hat{C}_\ell^{TV_i}$ and $\hat{C}_\ell^{TV_j}$ for bins
$i$ and $j$ at the same multipole $\ell$ is, under the Gaussian approximation,
\begin{equation}
\mathrm{Cov}\!\left[\hat{C}_\ell^{TV_i},\hat{C}_\ell^{TV_j}\right]
= \frac{1}{(2\ell+1)\fsky}
\Bigl[C_\ell^{TT}\,C_\ell^{V_iV_j}
+ C_\ell^{TV_i}\,C_\ell^{TV_j}\Bigr],
\label{eq:covariance}
\end{equation}
where $C_\ell^{TT}$ is the total (signal + noise) temperature power spectrum,
$C_\ell^{V_iV_j}$ is the cross-spectrum of the velocity-weighted convergence
fields between bins $i$ and $j$, and $\fsky = 0.2$ is the effective sky fraction, our estimate of the
realistically achievable overlap of the CMB, lensing, and spectroscopic
footprints (see Section~\ref{sec:snr_results}).
The first term dominates over the signal-squared term at all relevant $\ell$
for the observed-CMB cases, since the primary CMB variance is orders of
magnitude larger than the kSZ signal.
For the ideal cases we set $C_\ell^{TT} = C_\ell^{T_{\rm kSZ}T_{\rm kSZ}}$,
while for the observed-CMB cases $C_\ell^{TT}$ includes primary CMB,
instrument noise, and kSZ\rev{, $C_\ell^{TT}=B_\ell^2\,[C_\ell^{\rm CMB} +
C_\ell^{\rm kSZ} + N_\ell/1.5]$, which we evaluate analytically from the
inputs of Section~\ref{sec:T_obs} (the spectrum measured from the simulated
$T_{\rm obs}$ map agrees with it to within its scatter)}.
The covariance matrix $\mathrm{Cov}_\ell$ is an
$(N_{\rm bin}\times N_{\rm bin})$ matrix in bin index, computed from the full
$C_\ell^{V_iV_j}$ matrix, which we measure directly from the simulation
harmonic coefficients.

Beyond the Gaussian terms, the estimator here carries a connected contribution from
the long-wavelength radial velocity modes: because neighboring bins share the
same large-scale flows, the signal amplitude is modulated coherently across
$\sim 3$--$4$ adjacent bins and across all multipoles, in a manner analogous
to super-sample variance.
We observe this modulation directly in the simulation
at the
$\sim 10$--$20\%$ level per bin.
However, because the modulation is measured by and cancels against the calibrated
$A_{ij}$ within a single realization, it does not bias the estimator
and only degrades the
quoted SNR values very mildly (see Section~\ref{sec:snr_results}). 
In practice, this would be a negligible effect in real data.

\subsection{Fisher signal-to-noise}
\label{sec:fisher_snr}

The cumulative Fisher SNR is
\begin{equation}
\mathrm{SNR}^2 = \sum_\ell \mathbf{s}_\ell^T\,\mathrm{Cov}_\ell^{-1}\,\mathbf{s}_\ell,
\label{eq:fisher_snr}
\end{equation}
where $\mathbf{s}_\ell = (C_\ell^{TV_1,\rm sig}, \ldots,
C_\ell^{TV_{N_{\rm bin}},\rm sig})$ is the vector of signal power spectra.
For the observed cases the signal template is the beam-corrected ideal,
$B_\ell\,C_\ell^{TV,{\rm raw}}$, consistent with
Eq.~\eqref{eq:beam_correction}.
\rev{Two aspects of the template deserve comment. First, if the mean signal
$\mathbf{s}_\ell$ is the cross-spectrum measured in the \emph{same} simulation
realization as the covariance, its noise adds a positive plug-in bias,
$\mathrm{tr}[\mathrm{Cov}_\ell^{-1}\Sigma_\ell]$ with $\Sigma_\ell$ the
measurement covariance of the template, to the Fisher quadratic form. Summed
over all bins and multipoles this term is large and would inflate the quoted
significances. We subtract it multipole by multipole (an $\ell$-smoothed
template gives the same result), and for the ideal cases, where the correction
is not perturbative, we use the analytic template of
Eq.~\eqref{eq:master_discrete} instead. Second, the kSZ map used for the
validation of Section~\ref{sec:validation} contains only the gas of the tracer
volume; all forecast templates are instead measured against the map of the
full line of sight, $0.1<z<1.5$ (Appendix~\ref{sec:wings}).}
The sum starts at $\ell_{\rm min}=30$ and extends to $\lmax=8000$\rev{;
the contribution from $\ell<30$ is negligible}.

\section{Results}
\label{sec:results}

\subsection{Validation of the analytic model}
\label{sec:validation}

\begin{figure}[t]
\centering
\includegraphics[width=\linewidth]{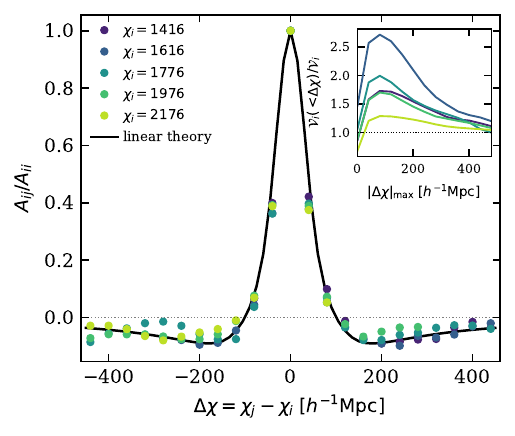}
\caption{
The measured velocity-correlation band matrix $A_{ij}/A_{ii}$ as a function of
radial bin separation $\Delta\chi = \chi_j - \chi_i$, for the five
representative bins (coloured points), compared to the bin-averaged
linear-theory correlation profile of Eqs.~\eqref{eq:xiv_structure} and
\eqref{eq:Psi_par} (black curve).
The positive core of width $\sim 40\;\mpc$ and the shallow negative tail
mandated by the sum rule of Eq.~\eqref{eq:sumrule} are both clearly detected.
The inset plot shows the cumulative velocity kernel
$\sum_{|j-i|\le n} \Delta\tau_j A_{ij}$ as a function of band width
$n$ for 5 representative bins, normalized by the total velocity kernel
over all bands,
illustrating the slow convergence driven by the negative tail and the
edge-bin enhancement.
}
\label{fig:velocity_kernel}
\end{figure}

We validate the master formula, Eq.~\eqref{eq:master_discrete}, against the
direct simulation measurement in the cleanest configuration
($T = T_{\rm kSZ}$, single source plane at $z_s = 1.9$; both measurement and
model carry the $1.6'$ beam and the pixel window).

\textit{Density leg.}
For each of the 24 bins, $j$, we measure $C_\ell^{\tau{\rm dm},j}$ from the projected
matter overdensity of the light-cone shells covering that bin (covered width
$\Delta\chi_{{\rm cov},j}\approx43$--$49\;\mpc$), with shot noise subtracted,
the pixel window deconvolved, and the gas transfer function of
Eq.~\eqref{eq:transfer} applied, so that under Limber
$\Delta\chi_{{\rm cov},j}\,C_\ell^{\tau{\rm dm},j}=\Pme(k_\ell)/\chi_j^2$.
The shells must be thick: a slab of width $\Delta\chi$ admits radial modes
$|k_\parallel|\lesssim2\pi/\Delta\chi$, so its auto-spectrum averages the
falling $\Pme$ over larger total $k$ and is biased low until
$\ell\gg2\pi\chi/\Delta\chi$: a bias absent from the measurement itself,
whose density leg (the cross with $\kappa$) selects $k_\parallel\approx0$
through the radially broad lensing kernel.
For our $\sim45\;\mpc$ stacks this bias is below $1\%$ at $\ell\gtrsim10^3$,
whereas single-step ($\sim6\;\mpc$) shells remain several percent low even at
$\ell\sim5000$.
Finally, because the spectra are measured per bin from the same realization
the kSZ is built from, the density-leg sample variance appears on both sides of
the comparison and largely cancels: the residuals of the theory prediction reflect modeling simplifications.

\textit{Velocity leg.}
The kernel is calibrated via the band matrix $A_{ij}$
(Eq.~\ref{eq:Aij_def}) over all 24 bins, with per-bin optical depths
$\Delta\tau_j$ and lensing kernels $W_\kappa(\chi_j)$. We compute the linear
theory prediction of the band matrix $A_{ij}$ and compare it with the simulation
result. The structure underlying the velocity kernel is shown directly in
Fig.~\ref{fig:velocity_kernel}, which plots the measured band matrix
$A_{ij}/A_{ii}$ as a function of radial bin separation: its positive core and
shallow negative tail reproduce the linear-theory correlation profile of
Eq.~\eqref{eq:Psi_par}, and the inset shows how the kernel $\mathcal{V}_i$
converges only once the negative tail is summed, which is the origin of the enhanced
sensitivity of the edge bins.

\begin{figure}[t]
\centering
\includegraphics[width=\linewidth]{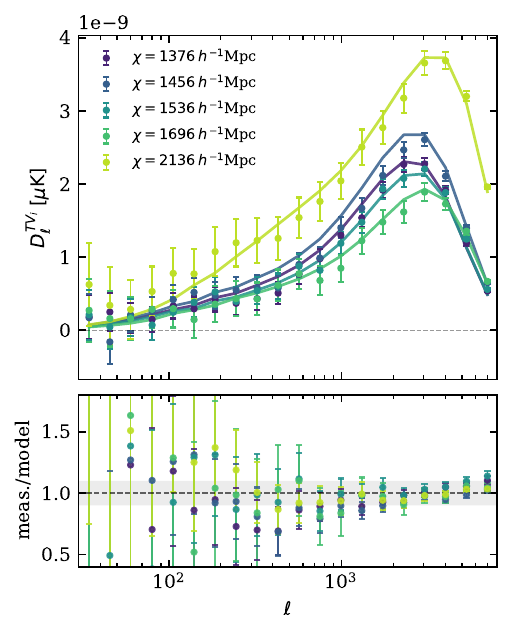}
\caption{
Validation of the analytic model (Eq.~\ref{eq:master_discrete}) against the
direct simulation measurement for five representative radial bins
($T = T_{\rm kSZ}$, source plane at $z_s=1.9$; both measurement and model
convolved with the $1.6'$ beam).
\emph{Top}: band-averaged $D_\ell^{TV_i}$; points with Gaussian error bars
show the measurement, solid curves the model with the velocity kernel
calibrated via the measured $A_{ij}$ and the density leg measured from the
stacked matter shells of each bin, with no free parameters.
\emph{Bottom}: ratio of measurement to model.
Agreement is at the $\lesssim5\%$ level (median over interior bins) for
$\ell \gtrsim 10^3$. 
}
\label{fig:theory_validation}
\end{figure}

Figure~\ref{fig:theory_validation} shows the comparison for five representative interior bins, demonstrating that the analytic model reproduces the measured signal remarkably well, with agreement at the $\lesssim 5\%$ level for $\ell\gtrsim500$. This agreement is not limited to these representative examples: in Appendix~\ref{app:pipeline}, we quantify the measurement-to-model ratio in the peak band, $\ell\in[1500,4000]$, for all 24 radial bins. The remaining differences arise from several simplifying approximations in the model, including fluctuations in the velocity and density legs, the use of relatively thick velocity bins rather than the mass-weighted velocities entering the kSZ signal, and other approximations that enable the model to retain its simple analytic form.

\begin{figure*}[t]
\centering
\includegraphics[width=\linewidth]{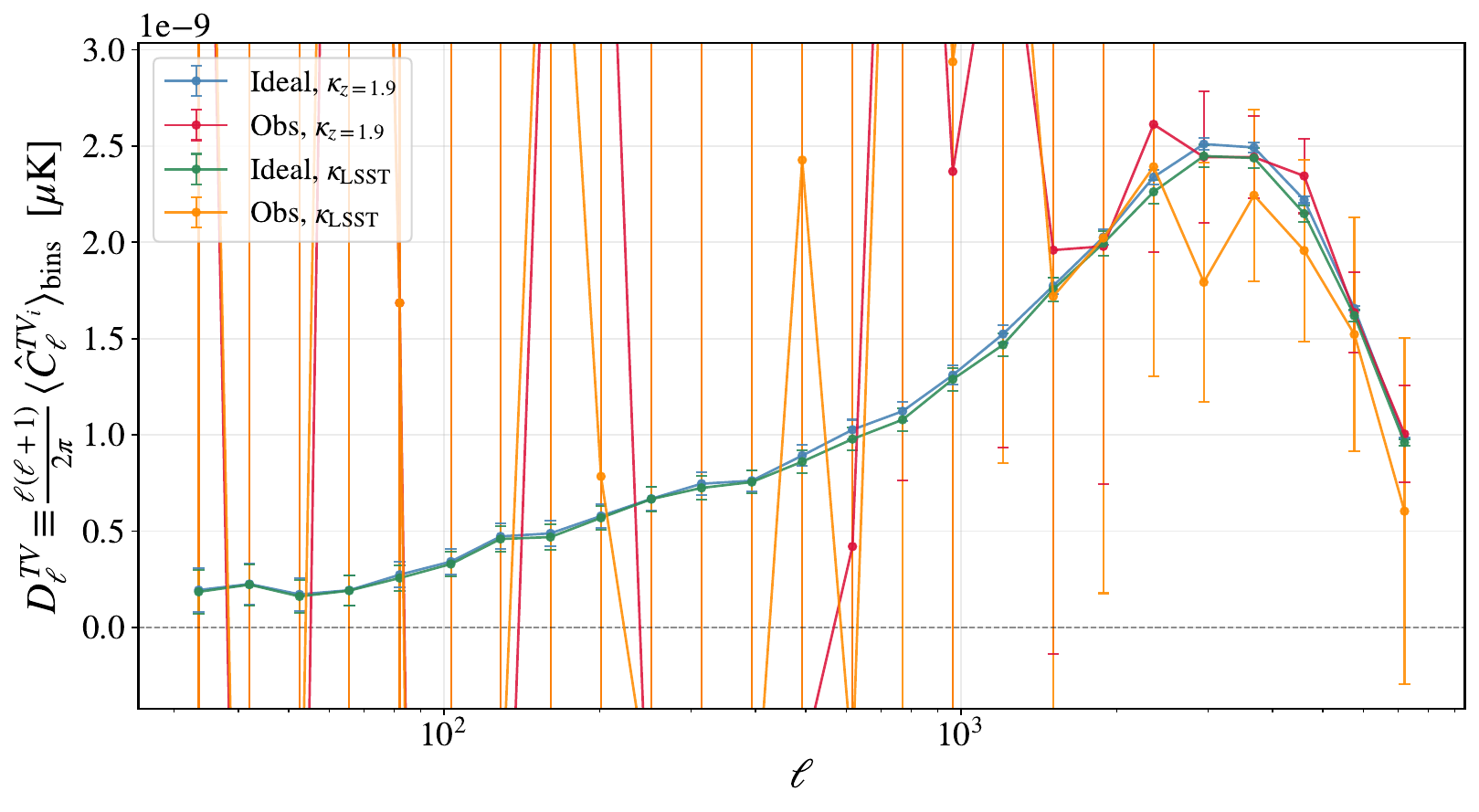}
\caption{
Band-averaged $D_\ell^{TV} \equiv \frac{\ell(\ell+1)}{2\pi}
\langle\hat{C}_\ell^{TV_i}\rangle_{\rm bins}$ for the four cases, averaged over
24 radial bins.
Ideal cases (beam-corrected ideal kSZ $\times$ beam):
Case~1a uses $\kappa_{z=1.9}$ (blue); Case~2a uses $\kappa_{\rm LSST}$ (green).
Observed cases:
Case~1b (red) is the $\kappa_{z=1.9}$ case with realistic CMB and noise;
Case~2b (orange) is the LSST case with realistic CMB and LSST shape noise.
All observed data points are consistent with the beam-corrected ideal within
their (large, CMB-dominated) error bars, confirming the estimator is unbiased.
The small difference between the  
$\kappa_{z=1.9}$ and $\kappa_{\rm LSST}$ case is due to the different lensing 
efficiencies.
}
\label{fig:bias_check}
\end{figure*}

\subsection{Unbiased estimator}
\label{sec:bias_check}

A fundamental consistency test is that the observed cross-spectrum
$\hat{C}_\ell^{TV}$ be an unbiased estimator of $B_\ell\,C_\ell^{TV,\rm raw}$
in the presence of the primary CMB and instrument noise.
We note that the noise properties of the reconstructed velocity field, convergence map 
and CMB map are uncorrelated and therefore they all cancel in the ensemble average,
leading us to expect that indeed the measured signal should be unbiased
relative to the theoretical signal. We test this explicitly in this section.
We note that since our test is on simulations, this requires all 
noise (stochastic) fields to be generated with independent random
seeds, as discussed in Appendix~\ref{app:pipeline}.

Figure~\ref{fig:bias_check} shows the band-averaged $D_\ell^{TV}$ for all four
cases, averaged over the 24 radial bins. We expect the ideal and observed cases
to match each other in pairs: i.e., observed LSST should match ideal LSST, and
same for the two $z_{\rm s} = 1.9$ cases.
For the ideal cases the beam-corrected spectra are shown as solid curves with
error bars estimated from the Gaussian covariance.
For the two observed cases the measured $D_\ell^{TV}$ is also estimated
from the Gaussian covariance, but this time it is dominated by the primary
CMB variance in $C_\ell^{TT}$, which inflates the per-bin error bars by
a factor of ${\sim}10$ relative to the CMB-free ones, and makes it close to impossible
to measure the signal at $\ell \lesssim 1000$.
The signal remains statistically consistent with the beam-corrected 
curves within the error bars at every $\ell$ band,
confirming that the estimator is unbiased (Section~\ref{sec:bias_check}). Furthermore, we check that the brute-force inferred signal-to-noise from this measurement on simulations is in very agreement with the signal-to-noise computed via our theoretical model Gaussian forecast.

\subsection{Per-bin signal structure}
\label{sec:per_bin}

Having validated the model and checked that the estimator is unbiased,
we examine the estimated signal structure
$D_\ell^{TV_i} \equiv \ell(\ell+1)/(2\pi)\,\hat{C}_\ell^{TV_i}$ for individual
bins under ideal and realistic observing conditions.

\begin{figure*}[t]
\centering
\includegraphics[width=\linewidth]{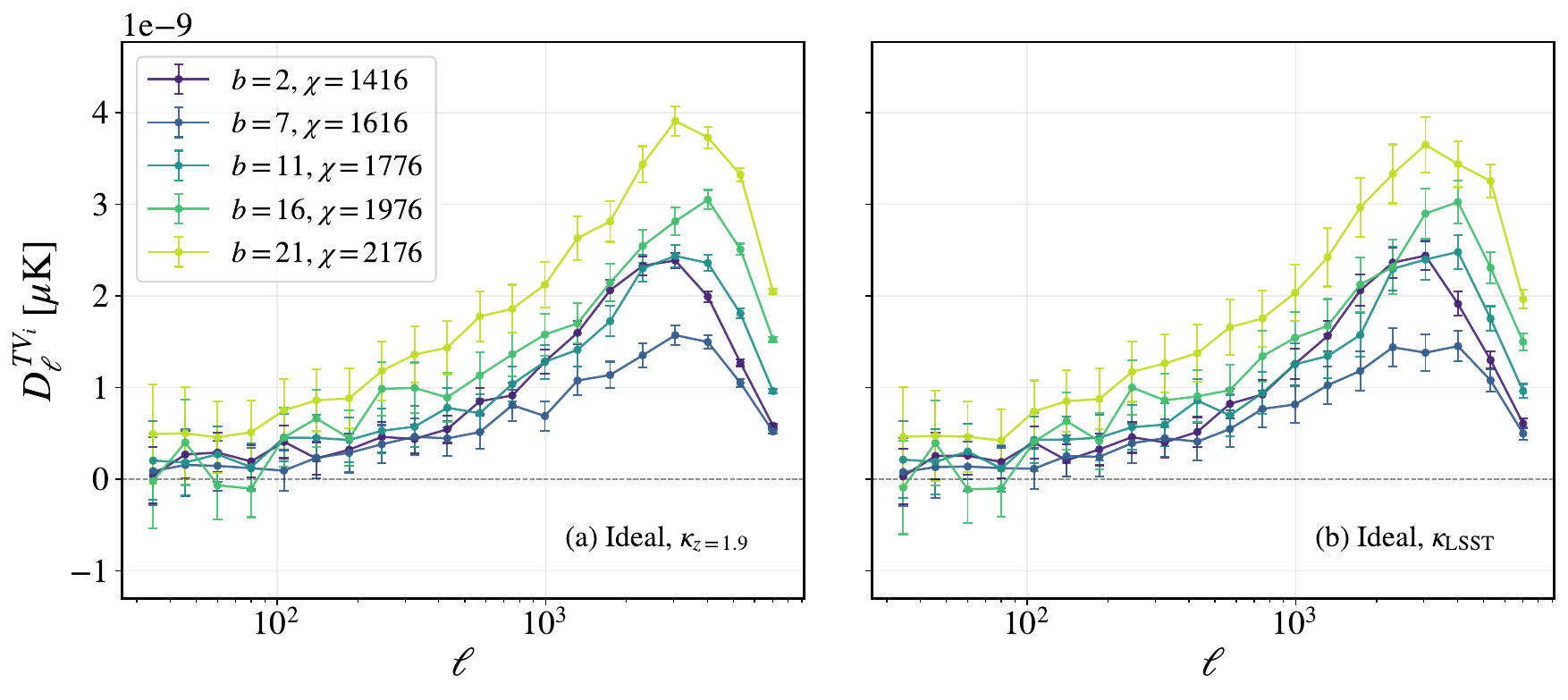}
\caption{
Band-averaged $D_\ell^{TV_i} \equiv \ell(\ell+1)/(2\pi)\,\hat{C}_\ell^{TV_i}$
for five representative radial bins (coloured curves) at four observational
settings.
\emph{Top row}: source plane at $z_s = 1.9$.
\emph{Bottom row}: \rev{LSST-like high-redshift source selection
($z_{\rm ph}\in[1.5,2.6]$, $\langle z\rangle\approx1.9$,
$n_\mathrm{eff}\approx4.6\,\mathrm{arcmin}^{-2}$, $\sigma_e=0.26$;
Section~\ref{sec:kappa_lsst})}.
\emph{Left column}: ideal case, $T = B_\ell T_\mathrm{kSZ}$ only.
\emph{Right column}: realistic observed CMB (beam FWHM$\,=1.6'$, Simons
Observatory-like ILC noise, primary CMB variance included).
Central values are the beam-corrected ideal signal; error bars follow
Eq.~\eqref{eq:per_bin_sigma} and are clipped at the panel boundaries.
The signal is detected per-bin in the ideal cases; in the observed cases the
CMB-dominated noise floor suppresses per-bin significance, motivating the
multi-bin Fisher combination. The small difference between the corresponding 
curves in the two panels are due to the different lensing efficiencies.
}
\label{fig:bins_comparison}
\end{figure*}

Figure~\ref{fig:bins_comparison} shows $D_\ell^{TV_i}$ for the five
representative radial bins.
The four panels correspond to the two ideal cases (i.e. without CMB noise) defined in
Section~\ref{sec:snr_results}: ideal $\kappa_{z=1.9}$ (left) and ideal LSST (right).
In each panel the central values are the beam-corrected raw signal
$B_\ell\,C_\ell^{TV_i,\mathrm{raw}}$, and the error bars represent the
$1\sigma$ Gaussian uncertainty on a single $\ell$-band measurement:
\begin{equation}
\sigma_{D_\ell^{TV_i}} = \frac{\ell(\ell+1)}{2\pi}
\sqrt{\frac{C_\ell^{TT}\,C_\ell^{V_i V_i} + \bigl(C_\ell^{TV_i,\mathrm{sig}}\bigr)^2}
          {(2\ell+1)\,\fsky}},
\label{eq:per_bin_sigma}
\end{equation}
where $C_\ell^{TV_i,\mathrm{sig}} = B_\ell\,C_\ell^{TV_i,\mathrm{raw}}$ for the
observed cases and $C_\ell^{TV_i,\mathrm{ideal}}$ for the ideal cases.
Error bars that exceed the panel range are clipped at the panel edge.

Several features are apparent.
The signal peaks at $\ell \sim$ a few thousand and is suppressed at
higher $\ell$ by the instrumental beam, consistent with
Eq.~\eqref{eq:beam_correction}.
In these CMB-free measurements, the signal is clearly detected across
the full multipole range for all five bins: the error bars are narrow because
$C_\ell^{TT}$ is the small kSZ-only auto-spectrum.

The amplitude difference between the ideal $\kappa_{z=1.9}$ and ideal LSST
cases is quantitatively accounted for by the source-averaged lensing kernel,
Eq.~\eqref{eq:Wkappa_eff}: the LSST \rev{high-redshift source selection (Eq.~\ref{eq:dndz}, with
a mean redshift of $\langle z \rangle \approx 1.9$)}
has a slightly lower mean lensing efficiency within the foreground slice $z=0.5$--$1$
than the single plane at $z_s=1.9$, reducing the signal by ${\lesssim}2\%$.
The relative ordering of bins is preserved between the two source
configurations.

\subsection{Covariance structure between radial bins}
\label{sec:covariance_structure}

A distinctive feature of the shear--kSZ estimator is that the fields $V_i$ and $V_j$
for different radial bins $i \ne j$ share the same lensing convergence
map $\kappa(\nhat)$.
This common factor induces non-trivial off-diagonal covariance between bins at
the same multipole $\ell$,
Eq.~\eqref{eq:covariance}.
The cross-spectrum $C_\ell^{V_i V_j}$ between bins $i$ and $j$ does not vanish
because
$\langle V_i(\nhat)\,V_j(\nhat')\rangle \supset
\langle\vtil_i\,\vtil_j\rangle\,\langle\kappa\,\kappa\rangle$,
which is non-zero both through the velocity correlation between neighboring
bins (the band structure of Fig.~\ref{fig:velocity_kernel}) and through the
shared $\kappa$ power.

We quantify this coupling with the correlation coefficient
\begin{equation}
\rho_{ij}(\ell) \equiv
\frac{\mathrm{Cov}\!\left[\hat{C}_\ell^{TV_i},\hat{C}_\ell^{TV_j}\right]}
{\sqrt{\mathrm{Cov}\!\left[\hat{C}_\ell^{TV_i},\hat{C}_\ell^{TV_i}\right]\,
\mathrm{Cov}\!\left[\hat{C}_\ell^{TV_j},\hat{C}_\ell^{TV_j}\right]}},
\label{eq:rho_def}
\end{equation}
with the covariance given by Eq.~\eqref{eq:covariance}.

\begin{figure*}[t]
\centering
\includegraphics[width=\linewidth]{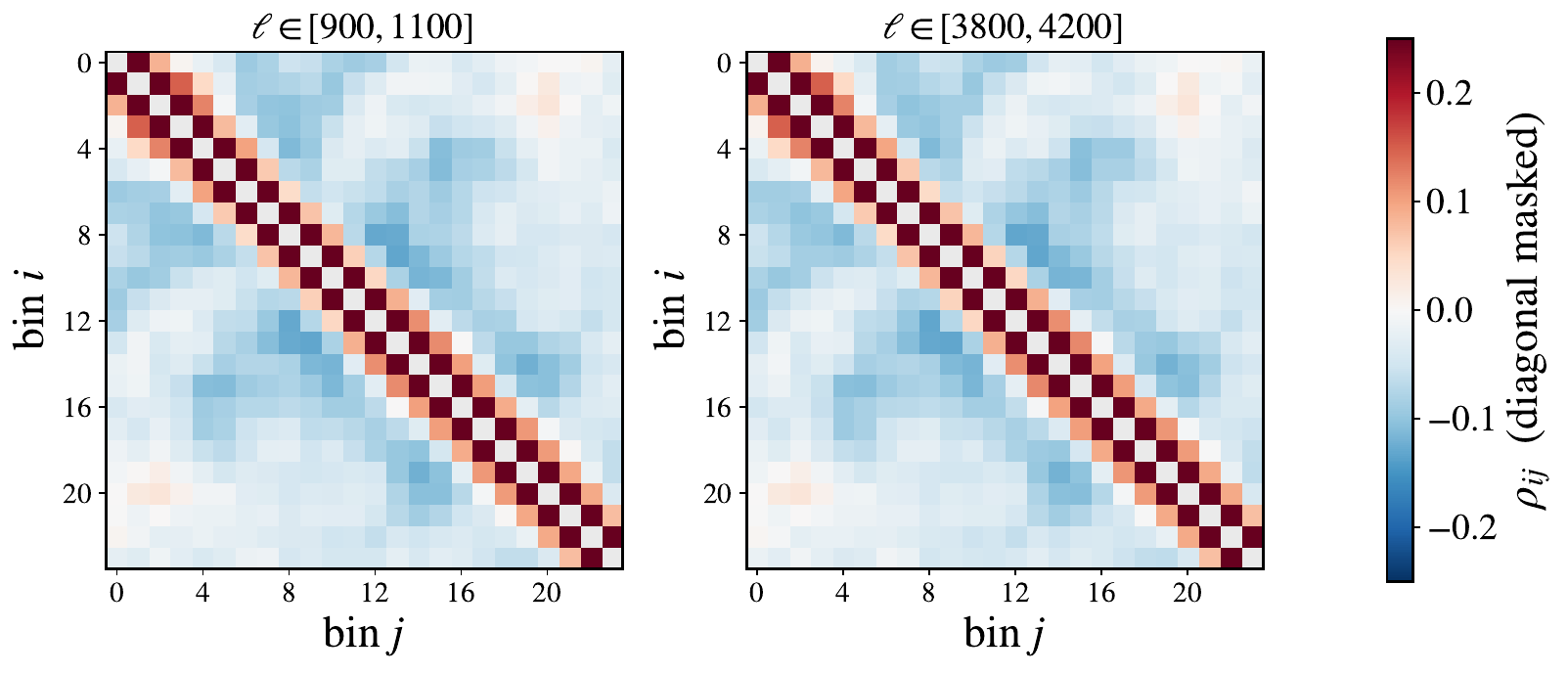}
\caption{
Inter-bin covariance structure of the estimator (ideal case, $T=T_{\rm kSZ}$,
$\kappa_{z=1.9}$; the LSST case is qualitatively similar).
Band-averaged correlation matrices $\rho_{ij}$
(Eq.~\eqref{eq:rho_def}) for all 24 radial bins in two multipole bands; the
diagonal ($\rho_{ii}=1$) is subtracted.
The covariance inherits the radial structure of the line-of-sight velocity
correlation function: a positive nearest-neighbour band and a shallow
negative lobe at $\Delta\chi\sim150$--$250\;\mpc$, because at Gaussian order
the bins couple only through the velocity cross-spectrum
$C_\ell^{\tilde v_i\tilde v_j}$; the shared $\kappa$ field alone induces no
inter-bin covariance.}
\label{fig:covariance}
\end{figure*}

Figure~\ref{fig:covariance} shows the band-averaged correlation matrices for
all 24 bins in two representative multipole bands, together with the mean
correlation as a function of radial bin separation.
The off-diagonal correlations are weak, $|\rho_{ij}|\lesssim0.15$, and
confined to a narrow band around the diagonal.
This structure is dictated by the velocity field: at Gaussian order,
$C_\ell^{V_iV_j} = \int\!\frac{d^2L}{(2\pi)^2}\,
C_L^{\tilde v_i\tilde v_j}\,C^{\kappa\kappa}_{|\boldsymbol\ell-\mathbf L|}$,
so bins couple only to the extent that their velocity templates correlate;
the fact that all bins multiply the same $\kappa(\nhat)$ field induces, by
itself, no covariance between them.
Accordingly, $\rho_{ij}$ inherits the radial profile of the line-of-sight
velocity correlation function established in
Section~\ref{sec:theory_velkernel}: a positive core for adjacent bins and a
shallow negative lobe at separations of $\sim150$--$250\;\mpc$.
The same velocity correlation function, encoded in the calibration matrix
$A_{ij}$, thus controls both the signal kernel of
Eq.~\eqref{eq:master_discrete} and the inter-bin covariance.
This serves as a nontrivial internal consistency check.
For the Fisher computation we invert the full
$(N_{\rm bin}\times N_{\rm bin})$ covariance at each $\ell$; given the weak
correlations, the diagonal approximation changes the total SNR by only less than a percent.
As noted in Section~\ref{sec:covariance}, the coherent velocity-amplitude
modulation 
constitutes an additional non-Gaussian
contribution, fully correlated across multipoles, that
Eq.~\eqref{eq:covariance} does not capture; the 24-bin scatter about the
smooth radial trend provides a direct estimate of its magnitude.
The per-bin measurement uncertainties underlying these error bars are shown as a function of scale in Appendix~\ref{app:pipeline} (Fig.~\ref{fig:sigma_vs_ell}).

For the Fisher computation we invert the full
$(N_\mathrm{bin}\times N_\mathrm{bin})$ covariance matrix $\mathrm{Cov}_\ell$
at each $\ell$ (Eq.~\eqref{eq:fisher_snr}), which accounts correctly for all
off-diagonal Gaussian correlations.
Ignoring these correlations and using only the diagonal
(Eq.~\eqref{eq:per_bin_sigma}) overestimates the SNR by
${\sim}15$--$20\%$ at low $\ell$ where $\rho_{ij}$ is largest.
As noted in Section~\ref{sec:covariance}, the coherent velocity-amplitude
modulation constitutes an additional,
non-Gaussian, positively correlated contribution across bins and multipoles
that Eq.~\eqref{eq:covariance} does not capture; the 24-bin scatter about the
smooth radial trend provides a direct estimate of its magnitude, but the change
to the SNR is only about 0.2\%.

\begin{revblock}
\subsection{Realistic velocity reconstruction and photometric redshifts}
\label{sec:recon}

\begin{figure*}[t]
\centering
\includegraphics[width=0.95\linewidth]{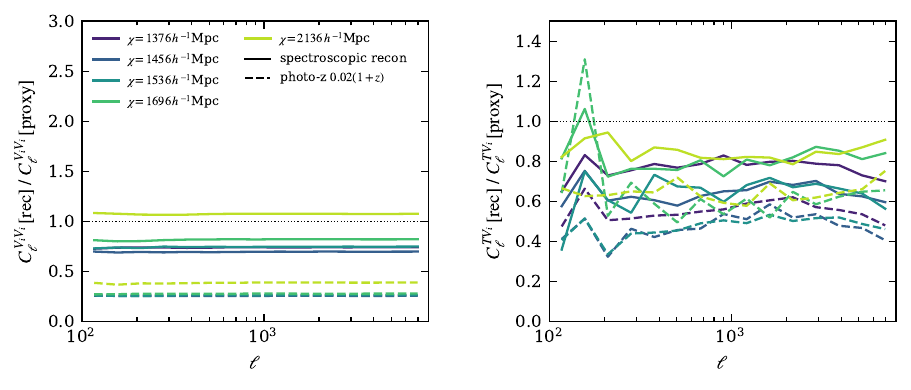}
\caption{
Reconstructed versus smoothed-true-velocity (proxy) templates for five
representative shells. \emph{Left}: ratio of the template auto-spectra
$C_\ell^{V_iV_i}$, which enter the covariance, for the spectroscopic
reconstruction (solid) and for the reconstruction with photometric-redshift
scatter $\sigma_z=0.02(1+z)$ (dashed). \emph{Right}: ratio of the signal
templates $C_\ell^{TV_i}$, i.e.\ the response. With spectroscopic redshifts the
response is $0.7$--$0.8$ of the proxy and the template variance $0.7$--$1.1$.
The photometric scatter halves the response but, acting as a radial low-pass
filter, also removes most of the small-scale reconstruction noise.}
\label{fig:recon_noise}
\end{figure*}

The proxy templates of Section~\ref{sec:vfield} capture the mean response of a
reconstructed velocity field, but not necessarily its noise. We therefore
repeat the case-2 forecast with templates built from velocities that
\texttt{pyrecon} reconstructs from the observed (redshift-space) halo positions,
once with spectroscopic redshifts and once with a Gaussian photometric-redshift
scatter $\sigma_z=0.02(1+z)$, typical of LSST red galaxies. Both runs share the
random catalog, the mesh and the object identifiers, so they differ only by the
added scatter. At the catalog level, the reconstructed line-of-sight velocities
correlate with the true ones at $r=0.65$ (spectroscopic) and $r=0.32$
(photometric), similar to the values in stacked kSZ analyses.

Figure~\ref{fig:recon_noise} compares the two sets of templates through the
quantities that enter the forecast. For the spectroscopic reconstruction, the
stochastic reconstruction noise is comparable to the smoothing losses, and the
SNR drops by a quarter relative to the proxy for the kSZ of the tracer volume,
and by about a half once the gas along the full line of sight is included
(Appendix~\ref{sec:wings}). The photometric scatter lowers the response further,
but it also acts as a radial low-pass filter that removes most of the
small-scale reconstruction noise, so the information per shell barely changes.
\emph{Photometric redshifts therefore cost only $\approx5$--$10\%$ of the
significance at fixed sky area} (cases 2c and 2d of Table~\ref{tab:snr}), even
though they halve the correlation with the true velocities. Halving the shell
width, or binning the reconstructed halos by their true rather than observed
distance, leaves the SNR unchanged: the information of the estimator saturates
at the radial coherence length of the velocity field, $\gtrsim100\;\mpc$, not
at the shell width. A photometric survey such as LSST can therefore supply its
own velocity templates over its full footprint, which we use in
Section~\ref{sec:snr_results}.
\end{revblock}

\subsection{Signal-to-noise ratio}
\label{sec:snr_results}

\begin{figure*}[t]
\centering
\includegraphics[width=\linewidth]{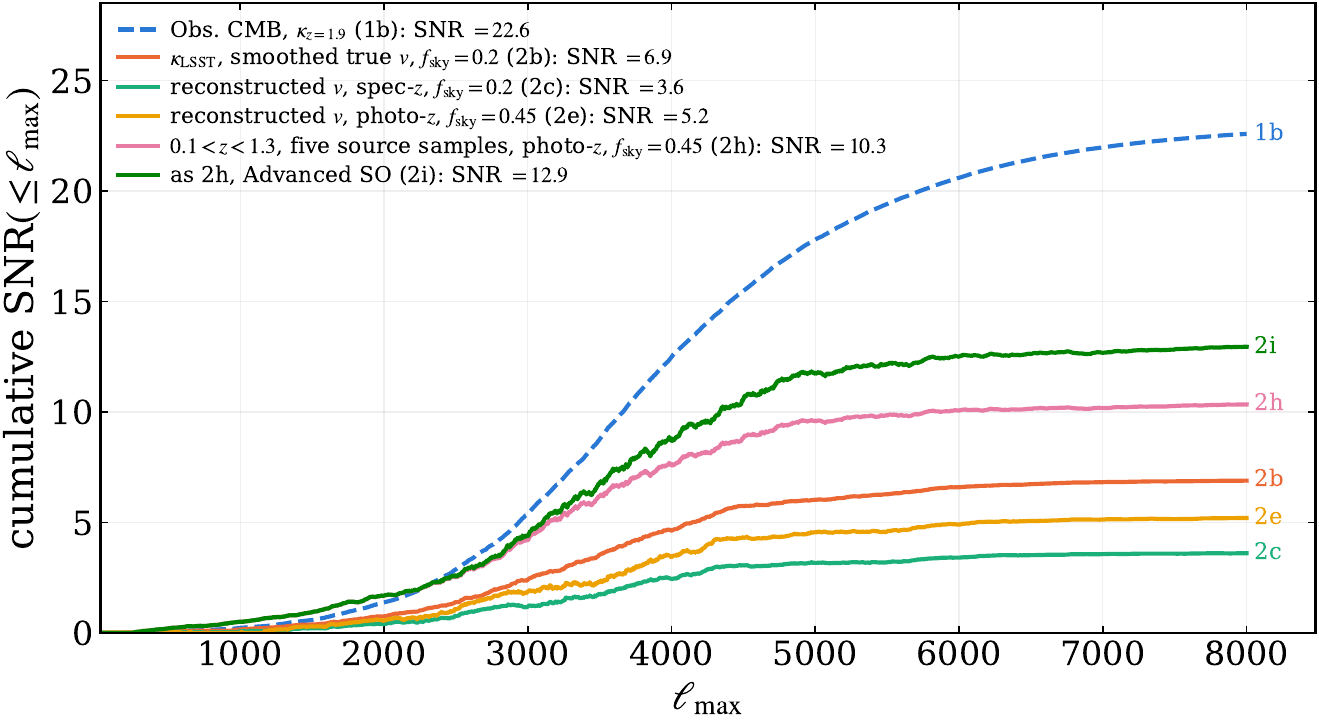}
\caption{
Cumulative Fisher signal-to-noise ratio as a function of $\ell_{\rm max}$
for the \rev{seven} shear--kSZ forecast cases \rev{of Table~\ref{tab:snr}}.
Solid curves denote ideal configurations (kSZ-only $C_\ell^{TT}$);
\rev{the other line styles} denote realistic observed configurations.
Blue and red: source plane at $z_s=1.9$.
Green and orange: LSST-like high-$z$ source sample
(Section~\ref{sec:kappa_lsst}) \rev{with the smoothed-true-velocity templates;
purple, grey and black: LSST sources with the reconstructed velocity
templates of Section~\ref{sec:recon} (spectroscopic; photometric; photometric
over $\fsky=0.45$)}.
The SNR values in the legend correspond to the total integrated Fisher SNR.
}
\label{fig:snr}
\end{figure*}

Figure~\ref{fig:snr} shows the cumulative Fisher SNR as a function of
$\ell_{\rm max}$ for representative cases, and Table~\ref{tab:snr} lists the
totals for all of them\rev{, including the extended-volume cases 2f--2i of
Section~\ref{sec:extvol}}.

\begin{table}[h]
\centering
\begin{tabular}{llc}
\hline\hline
Case & Description & SNR \\
\hline
1a & Ideal, $\kappa_{z=1.9}$ & \rev{$224$} \\
1b & Obs.\ CMB, $\kappa_{z=1.9}$ & \rev{$22.6$} \\
2a & Ideal, $\kappa_{\rm LSST}$ & \rev{$73$} \\
2b & Obs.\ CMB, $\kappa_{\rm LSST}$, smoothed true $v$ & \rev{$6.9$} \\
\rev{2c} & \rev{Obs.\ CMB, $\kappa_{\rm LSST}$, reconstructed $v$ (spec-$z$)} & \rev{$3.6$} \\
\rev{2d} & \rev{Obs.\ CMB, $\kappa_{\rm LSST}$, reconstructed $v$ (photo-$z$)} & \rev{$3.5$} \\
\rev{2e} & \rev{as 2d, $f_{\rm sky}=0.45$} & \rev{$5.2$} \\
\rev{2f} & \rev{as 2c, $0.3<z<1.1$, two source samples} & \rev{$6.9$} \\
\rev{2g} & \rev{as 2f, photo-$z$, $f_{\rm sky}=0.45$} & \rev{$9.1$} \\
\rev{2h} & \rev{as 2g, $0.1<z<1.3$, five source samples} & \rev{$10.3$} \\
\rev{2i} & \rev{as 2h, Advanced SO noise} & \rev{$13.0$} \\
\hline\hline
\end{tabular}
\caption{Total Fisher signal-to-noise ratios for the shear--kSZ forecast cases,
summed over $\ell = \rev{30}$--$8000$ with $f_{\rm sky}=0.2$ \rev{(0.45 for cases
2e and 2g--2i)} under the Gaussian covariance of Eq.~\eqref{eq:covariance}. The
``ideal'' cases use only the kSZ contribution to $C_\ell^{TT}$ \rev{(the shape
noise of the LSST sources still enters through $C_\ell^{V_iV_j}$)}; the
``observed'' cases include primary CMB, Simons Observatory-like ILC noise
(FWHM\,$=1.6'$) and LSST shape noise. \rev{All cases use the kSZ map of the full
line of sight, $0.1<z<1.5$ (Appendix~\ref{sec:wings}). Cases 2c--2i use
velocities reconstructed with \texttt{pyrecon} (Section~\ref{sec:recon}), with
a photometric-redshift scatter $\sigma_z=0.02(1+z)$ for 2d, 2e and 2g--2i.
Cases 2f and 2g extend the tracers to $0.3<z<1.1$ and add a second source
sample for the shells at $z<0.75$; case 2h uses tracers at $0.1<z<1.3$ and five
source samples, each shell paired only with the samples that lie entirely
behind it (Section~\ref{sec:extvol}); case 2i repeats 2h with the Advanced SO
noise.}
}
\label{tab:snr}
\end{table}

The SNR is dominated by multipoles $\ell \sim 10^3$--$5\times10^3$, where the
kSZ signal is competitive with the Silk-damped primary CMB and the beam has
not yet strongly suppressed the signal.
The primary CMB variance reduces the SNR by a factor of $\approx 10$ relative
to the ideal case for both the $z=1.9$ and LSST scenarios.
The LSST configuration achieves \rev{$\approx 1/3$} of the SNR of the $z=1.9$
configuration in the ideal cases, with the reduction attributable to the
\rev{shape noise of the sparse high-redshift source sample (which enters
$C_\ell^{V_iV_j}$ even when $T=T_{\rm kSZ}$) and to the} lower
source-averaged lensing kernel (Eq.~\ref{eq:Wkappa_eff}).

\rev{In the realistic scenarios, SO-like CMB data combined with LSST lensing
over $f_{\rm sky}=0.2$ give $\mathrm{SNR}\approx7$ with the
smoothed-true-velocity proxy and $\approx3.5$ with reconstructed velocities
(cases 2b--2d). Because photometric redshifts cost little
(Section~\ref{sec:recon}), a photometric survey can provide its own velocity
templates. With LSST red galaxies as tracers over the full LSST--SO overlap,
$f_{\rm sky}=0.45$, the significance rises to $\approx5$ (case 2e), to
$\approx9$ with a larger tracer volume and a second source sample (case 2g), to
$\approx10$ with tracers at $0.1<z<1.3$ and five source samples (case 2h,
Section~\ref{sec:extvol}), and to $\approx13$ with Advanced SO (case 2i). The
signal is therefore detectable at $\gtrsim10\sigma$ with the upcoming surveys.}

\rev{Interestingly, we find that the CMB temperature map dominates the noise:
primary CMB and instrument noise contribute comparably at the multipoles that
carry the signal. The forecast therefore depends only mildly on the depth of
the lensing survey: LSST Y1-like (Y3-like) shape noise would lower the
significance by $\approx35\%$ ($\approx15\%$). Conversely, Advanced SO, with
three times less noise power than nominal SO, raises the SNR by only
$\approx50\%$, because the primary CMB variance cannot be reduced; a
noise-free CMB map would give $\mathrm{SNR}\approx37$ for case 2b. With the
nominal rather than the enhanced SO sensitivity, all observed-case values drop
by $14\%$.}

One assumption deserves explicit caution. Since $\mathrm{SNR}\propto\sqrt{\fsky}$,
the case-2 values assume that the CMB, lensing and spectroscopic footprints
share $20\%$ of the sky, which the DESI$\times$SO$\times$LSST overlap allows.
Other combinations may overlap more with Simons Observatory, for example
\textit{Euclid}, 4MOST or a Spec-S5--class survey with southern coverage, or
photometric tracers with a correspondingly noisier velocity reconstruction.
Section~\ref{sec:extvol} explores several tracer volumes and source samples for
LSST, but we have not tested every combination, and better choices may well
exist.

\begin{revblock}\subsection{Extending the tracer volume and combining source samples}
\label{sec:extvol}

The forecasts above use tracers at $0.5<z<1$ and a single high-redshift source
sample. A photometric survey such as LSST provides velocity tracers over a much
larger volume, and different source samples can be paired with different
shells, as long as a source sample has no support below the shell it is paired
with, which preserves the absence of intrinsic alignments and boost factors.
Two properties of the estimator make this configuration space easy to explore.

First, the radial resolution of the templates does not matter: merging the 24
shells of case~2b into a few shells of several hundred $\mpc$ changes the SNR
by less than $1\%$, and a single $960\;\mpc$ template keeps most of it. At the
multipoles that carry the signal, the estimator is therefore equivalent to
correlating the temperature map with a single velocity template, weighted along
the line of sight by $w_\ell=\taup W_\kappa\Pme(k_\ell)/\chi^2$, and its
information is the variance of the velocity field filtered by this slowly
varying weight. This is why the sum rule of Eq.~\eqref{eq:sumrule} matters
(Appendix~\ref{sec:wings}), and it agrees with the optimal shell width of order
$1\,h^{-1}$Gpc found by Ref.~\citep{2026arXiv260729091G}.

Second, this structure makes the forecast analytic to good accuracy.
Appendix~\ref{app:analytic} writes the Fisher information in terms of the
linear-theory velocity correlation functions, the Limber density leg, and the
source, noise and beam models of the simulation. Normalized once to the
measured full-line-of-sight value of case 2b, the analytic forecast reproduces
the simulated cases of Table~\ref{tab:snr} to $\lesssim10\%$, including the gain
from a second source sample, and we use it to explore tracer volumes and source
samples that we have not simulated.

Table~\ref{tab:extvol} summarizes the exploration. Extending the tracer volume
from $0.5<z<1$ to $0.1<z<1.1$ raises the significance by about a third, and
pairing the lower-redshift shells with lower-redshift source samples, which are
denser and lens those shells more efficiently, adds another third. Going beyond
$z\simeq1.3$ adds little, because only the sparse $z_{\rm ph}>2$ sources lie
entirely behind those shells. For LSST alone, with photometric reconstruction
over $\fsky=0.45$, tracers at $0.1<z<1.3$ and five source samples, the analytic
forecast gives $\mathrm{SNR}\approx11$, and $\approx14$ with Advanced SO. The
end-to-end simulation of the same configuration gives $10.3\pm0.9$ and
$13.0\pm1.2$ (cases 2h and 2i of Table~\ref{tab:snr}), where the error is the
scatter of the debiased plug-in estimate. The lower block of
Table~\ref{tab:extvol} lists further improvements: extending the tracers to
$z=1.5$, allowing a small source--lens overlap, better source photometric
redshifts, finer source binning, and the DESI spectroscopic reconstruction
where DESI overlaps. Each adds a few per cent; together they bring
LSST$\times$Advanced SO to $\approx16\sigma$. Delensing the temperature map
gains almost nothing, because at the relevant multipoles lensing adds less to
$C_\ell^{TT}$ than the unlensed damping tail and the instrument noise. The
largest remaining gain would come from the fidelity of the velocity
reconstruction: with noiseless velocities the same configurations would reach
$26$--$29\sigma$.

We note that combining several source samples in the Fisher matrix is
equivalent to the nulling transformation of Ref.~\citep{2026arXiv260729091G},
an invertible change of basis of the source bins, so these forecasts already
contain the localization of the lensing kernel that nulling provides. We leave
a full optimization of the source binning, and the treatment of the residual
source--lens overlap of the lower-redshift samples, to future work.

\begin{table*}[t]\color{black}
\centering
\begin{tabular}{lcccc}
\hline\hline
Tracers, source samples & \shortstack{$\fsky=0.2$\\spec-$z$ recon} & \shortstack{$\fsky=0.2$\\photo-$z$ recon} & \shortstack{$\fsky=0.45$\\photo-$z$ recon} & \shortstack{$\fsky=0.45$\\true $v$} \\
\hline
$0.5<z<1.0$, S$_{1.5\text{-}2.6}$ (case 2) & 4.2 & 3.7 & 5.6 & 10.4 \\
$0.3<z<1.1$, S$_{1.5\text{-}2.6}$ & 5.6 & 4.9 & 7.4 & 13.7 \\
$0.3<z<1.1$, + S$_{1.0\text{-}1.5}$ for $z<0.75$ & 6.6 & 5.9 & 8.8 & 16.3 \\
$0.1<z<1.1$, S$_{1.5\text{-}2.6}$ & 5.8 & 5.1 & 7.7 & 14.2 \\
$0.1<z<1.1$, four samples & 7.5 & 6.7 & 10.0 & 18.5 \\
$0.1<z<1.3$, S$_{1.5\text{-}2.0}$ + S$_{2.0\text{-}3.0}$ & 6.7 & 5.9 & 8.9 & 16.5 \\
$0.1<z<1.3$, five samples & 8.2 & 7.3 & 10.9 & 20.2 \\
$0.1<z<1.5$, five samples & 8.5 & 7.6 & 11.3 & 21.0 \\
\hline
$0.1<z<0.5$ (BGS-like), four samples & 3.4 & 3.0 & 4.5 & 8.3 \\
$1.0<z<1.5$ (ELG-like), S$_{1.5\text{-}2.0}$ + S$_{2.0\text{-}3.0}$ & 2.3 & 2.0 & 3.0 & 5.6 \\
\hline\hline
LSST alone, $\fsky=0.45$, photo-$z$ recon & SO & \shortstack{SO, + DESI\\spec-$z$ recon} & Advanced SO & \shortstack{Adv.\ SO, + DESI\\spec-$z$ recon} \\
\hline
$0.1<z<1.3$, five samples (as above) & 10.9 & 11.6 & 14.0 & 14.8 \\
\quad tracers to $z=1.5$ & 11.3 & 12.0 & 14.6 & 15.4 \\
\quad $5\%$ ($10\%$) source--lens overlap & 11.4 (11.5) & 12.1 (12.2) & 14.7 (14.8) & 15.6 (15.7) \\
\quad source photo-$z$ scatter $0.03(1+z)$ & 11.4 & 12.0 & 14.6 & 15.5 \\
\quad seven source samples & 11.1 & 11.7 & 14.3 & 15.1 \\
\quad delensing of $T$, $50\%$ ($80\%$) & 11.0 (11.1) & 11.7 (11.7) & 14.2 (14.4) & 15.1 (15.2) \\
\quad tracers to $z=1.5$ and $5\%$ overlap & 11.9 & 12.6 & 15.3 & 16.2 \\
\quad same, with source photo-$z$ $0.03(1+z)$ and $50\%$ delensing & 12.2 & 12.9 & 15.8 & 16.7 \\
\hline\hline
\end{tabular}
\caption{Analytic forecast of the signal-to-noise ratio
(Appendix~\ref{app:analytic}), normalized to the simulation, for the full line
of sight, as a function of the tracer volume and of the source samples.
S$_{a\text{-}b}$ denotes the LSST Y10 galaxies with $z_{\rm ph}\in[a,b]$;
``four samples'' are $[0.5,0.8]$, $[0.8,1.1]$, $[1.1,1.5]$ and $[1.5,2.6]$, and
``five samples'' split the last one into $[1.5,2.0]$ and $[2.0,3.0]$. Each
shell is paired only with the samples that have no support below it. Upper
block: DESI-like spectroscopic and LSST photometric reconstruction over
$\fsky=0.2$, photometric reconstruction over $\fsky=0.45$, and noiseless
velocities, all with the enhanced SO noise. Lower block: improvements applied
one at a time (last two rows: combined) to the best LSST-alone row, for SO and
Advanced SO, with and without the DESI spectroscopic reconstruction on the
$\fsky=0.2$ where DESI overlaps. The spectroscopic and the $\fsky=0.45$
photometric columns agree with the simulated cases 2c--2h of
Table~\ref{tab:snr} to $\lesssim10\%$, except for case 2c ($17\%$), whose
reconstruction volume is smaller.}
\label{tab:extvol}
\end{table*}
\end{revblock}

\section{Discussion and Conclusions}
\label{sec:discussion}

We have presented and validated shear--kSZ, a three-field cross-correlation of the
kSZ temperature, tomographically binned line-of-sight velocity templates, and
the weak-lensing convergence.
Its expectation value factorizes into a calibratable velocity
kernel multiplying the projected matter--electron cross-power spectrum
(Eq.~\ref{eq:master_discrete}), providing a direct window onto $\Pme(k)$ and,
through the exact species decomposition of
Eq.~\eqref{eq:suppression_exact}, onto the baryonic suppression of the matter
power spectrum.

Our key results are:
\begin{itemize}
\item The estimator's analytic model is \emph{validated end-to-end}: with the
velocity kernel calibrated from the tracer catalog (the band matrix
$A_{ij}$) and the density leg measured from thin matter shells, the prediction
matches the simulation measurement at the $\lesssim 5\%$ level over the
signal-dominated multipole range.
The calibration philosophy mirrors the velocity-reconstruction transfer function 
of standard stacked kSZ analyses and absorbs template normalization, smoothing
losses, tracer sampling, and the realization's long-wavelength velocity modes.
\item The estimator is \emph{unbiased}: with independent noise seeds, the
observed-case measurements are consistent with the beam-corrected signal within
the CMB-dominated error bars at all $\ell$ bands.
\item The estimator has \emph{a robust forecast SNR}: \rev{with reconstructed
velocities, the ionized gas along the full line of sight and the
same-realization template bias removed, the realistic LSST
scenarios yield $\mathrm{SNR}\approx4$--$7$ for a DESI-like sample over
$f_{\rm sky}=0.2$, $\approx10$ when LSST supplies its own velocity tracers at
$0.1<z<1.3$ over its full overlap with SO, and $\approx13$ with Advanced SO
(up to $\approx16$ with the improvements of Table~\ref{tab:extvol})}, under the Gaussian covariance.
\rev{Photometric redshifts lower the significance by only $\approx5$--$10\%$ at
fixed sky area, so that LSST can supply its own velocity templates over its
full footprint.}
\item The estimator is \emph{physically interpretable}: the direct
$\tau\times$matter cross-correlation in the simulation matches the ideal-case
estimator, and the line-of-sight velocity sum rule
(Eq.~\ref{eq:sumrule}) explains the tomographic design, the slow convergence
of the velocity kernel, the edge-bin behavior\rev{, and the reduction of the
signal by the ionized gas outside the tracer volume ($30$--$50\%$ depending on
the velocity templates), which we measure in the simulation with the kSZ of
the full line of sight}.
\end{itemize}

\subsection{Connection to cosmic shear and baryonic feedback}

One of the most pressing systematics for Stage-IV weak lensing surveys is the
uncertainty in the suppression of the non-linear matter power spectrum due to
baryonic feedback.
Current approaches include marginalisation over flexible baryon correction
models \citep{Mead2021, SchneiderTeyssier2015, Schneider2019}
or calibration on hydrodynamical simulations \citep{Chisari2019, vanDaalen2020}, but no direct
observational probe of $\Pme(k)$, which, via
Eq.~\eqref{eq:suppression_exact}, determines the suppression up to a
sub-percent backreaction term and a small, bounded electron auto-spectrum
correction, currently exists (see App.~\ref{app:suppression}).
shear--kSZ provides exactly this handle.
A joint analysis of shear--kSZ with cosmic shear from the same survey would allow
the baryonic suppression to be constrained internally, rather than
marginalized over.

\subsection{Connection to the stacked kSZ estimator}

The shear--kSZ estimator is the lensing-based analogue of the stacked kSZ
estimator.
Both exploit the same physical mechanism: the line-of-sight velocity of
foreground structure correlates with the kSZ temperature, and the density
field at the same position sets the cross-power.
The key distinction is the density proxy:
\begin{equation}
\underbrace{\langle v_{\rm rec}\,\delta_g, T_{\rm kSZ}\rangle
\propto P_{ge}}_{\text{stacked kSZ}}
\quad\longrightarrow\quad
\underbrace{\langle v_{\rm rec}\,\kappa, T_{\rm kSZ}\rangle
\propto \Pme}_{\text{shear--kSZ}}.
\label{eq:connection}
\end{equation}
The stacked kSZ measurement constrains $P^{ge}$ for a given tracer population
and requires combining information across tracers spanning a range of halo
masses to reconstruct the full matter suppression. The shear--kSZ estimator 
sidesteps this by using $\kappa$ as the unbiased matter proxy, at the
cost of the lensing kernel's broad radial support\rev{, which the tomographic
velocity binning only partly compensates (Appendix~\ref{sec:wings}),} and lensing shape noise.
\rev{Ref.~\citep{2026arXiv260729091G} proposes the same cross-correlation
with a two-index estimator (separate lensing and velocity bins) and a nulling
transformation of the source bins that localizes the lensing kernel; their
optimal velocity-shell width, of order $1\;h^{-1}{\rm Gpc}$, is consistent with our
finding that the radial information of the estimator saturates at the
coherence length of the velocity field rather than at the shell width.}

\subsection{Tracer samples and survey configurations}
\label{sec:tracers}

This forecast uses LRG-like halos at $z=0.5$--$1$, a natural choice given the
spectroscopic samples available from DESI \citep{2023AJ....165...58Z}.
The estimator is not restricted to LRGs, and the trade-offs for other
configurations follow directly from the structure of the signal and noise.
Moving to a lower-redshift tracer sample (e.g., BGS at $z\lesssim0.5$ with
sources at $z_s\sim1$) brings three penalties: a $\sim4\times$ smaller
comoving volume and hence fewer independent velocity and density modes, a
lower differential optical depth ($\taup\propto(1+z)^2$), and, most
importantly, the migration of fixed physical scales to lower multipoles where
the primary CMB variance per mode is an order of magnitude larger than in the
Silk-damped regime, though this is partially offset by reduced beam suppression, a larger
$\Pme/\chi^2$, more abundant background sources, and higher-fidelity velocity
reconstruction. \rev{With the analytic forecast of Appendix~\ref{app:analytic}
we find that a BGS-like volume ($0.1<z<0.5$) paired with the high-redshift
source sample alone yields $55\%$ less SNR than the $z=0.5$--$1$ configuration,
but only $20\%$ less when it is paired with lower-redshift source samples
(Table~\ref{tab:extvol}), and that the two volumes combined give more than the
quadrature sum of the two (the negatively correlated velocity noise of distant
shells is exploited by the Fisher combination); an ELG-like volume at
$1<z<1.5$ adds little with LSST sources, since only the sparse $z_{\rm ph}>2$
galaxies lie behind it.} The low-$z$
sample \rev{also} provides valuable systematic cross-checks and probes feedback at later
times.

We note that for the lensing convergence/shear catalogs, one can also adopt the CMB lensing reconstructed $\kappa$ map, which would yield high signal-to-noise using our estimator. There are several benefits associated with using CMB lensing maps: a) they can be used with higher-redshift foreground samples, behind which typically there are fewer sources and thus galaxy-shape lensing is not feasible; b) if constructed from the SO CMB LAT maps, they offer maximal overlap with the SO CMB temperature map; c) they do not suffer from various contaminants typical in galaxy-shape lensing such as intrinsic alignments, photometric redshift uncertainties, boost factors, etc. In practice, the main limitation of using $\kappa$ on very small scales, $L > 3000$, where a good fraction of our signal lies, is that at these scales, there are contributions from foregrounds, such as tSZ and CIB (although one can potentially apply various cleaning techniques to thwart that issue). However, Simons Observatory will provide unprecedentedly accurate $\kappa$ maps constructed using polarization-only, for which small-scale foreground contamination would not be a worry. We leave a more detailed study of CMB lensing application for our estimator for future work.

\begin{figure*}[t]
\centering
\includegraphics[width=\linewidth]{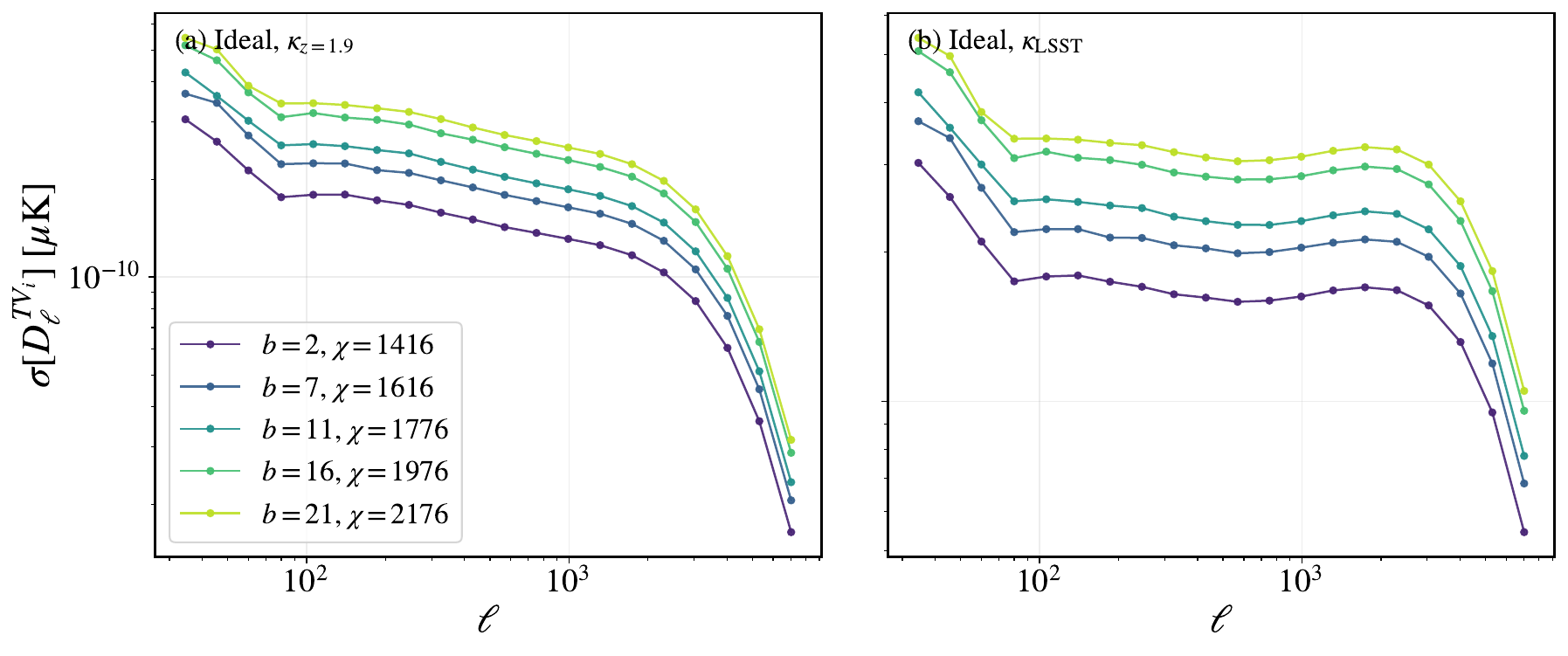}
\caption{
Per-band Gaussian uncertainty $\sigma[D_\ell^{TV_i}]$
(Eq.~\eqref{eq:per_bin_sigma}) for the five representative radial bins.
\emph{Left}: ideal case ($T=T_{\rm kSZ}$, $\kappa_{z=1.9}$);
\emph{right}: ideal LSST case ($T=T_{\rm kSZ}$, $\kappa_{\rm LSST}$ with shape
noise).
The steep low-$\ell$ decline reflects the growing number of modes per
logarithmic band ($\sim\ell^2\fsky$); the mid-$\ell$ plateau and high-$\ell$
drop trace the shape of $C_\ell^{TT}C_\ell^{V_iV_i}$.
Compared with the signal spectra of Fig.~\ref{fig:bins_comparison}, which
peak at $\ell\sim2$--$4\times10^3$, the per-band SNR is maximal in the same
range. No beam noise is applied to these plots (which would blow up the uncertainties on small scales).}
\label{fig:sigma_vs_ell}
\end{figure*}

Additionally, for the source sample we adopt a mock sample resembling the LSST Y10
golden sample \citep{DESC_SRD}. However, interestingly, we find that while shape noise degrades
the signal-to-noise somewhat, the main hit it takes is from the CMB primary 
and instrument ``noise''. \rev{If the three surveys share $20\%$ of the sky, we forecast
${\rm SNR}\approx4$--$7$ at full depth (Table~\ref{tab:snr}); LSST Y1 shape noise
lowers this by $\approx35\%$. Because photometric redshifts cost only
$\approx5$--$10\%$ of the significance (Section~\ref{sec:recon}), the most
promising configuration uses LSST for both the sources and the velocity tracers
over its full overlap with SO, $\fsky\approx0.45$: we forecast ${\rm SNR}\approx5$
for tracers at $0.5<z<1$ and $10$ for tracers at $0.1<z<1.3$ with five source
samples (Section~\ref{sec:extvol}).}
Other surveys such as \textit{Euclid} will be especially
well poised to complete the measurement due to both its larger overlap with DESI,
and the auspicious presence of both spectroscopic samples and (low-shape-noise)
shape measurements. Advanced Simons Observatory is projected to \rev{improve the 
signal-to-noise of this estimator by a further $\sim30\%$ relative to the enhanced SO sensitivity assumed here ($\sim50\%$ relative to nominal SO)}.

\subsection{Limitations and outlook}

The main simplifications of the present forecast are:
(i) the \rev{Smail-type parent distribution with an idealized Gaussian photo-$z$
window} for the LSST source distribution
(Eq.~\ref{eq:dndz}), an idealization of the true photometric distribution and
its calibration uncertainties;
(ii) \rev{the reconstruction of Section~\ref{sec:recon} is run on the full
halo sample of the simulation ($\bar n\simeq1.7\times10^{-3}\,h^3{\rm Mpc}^{-3}$),
denser than a DESI LRG sample, so its fidelity is representative of a dense
photometric sample rather than of DESI LRGs alone; and the ionized gas is
included out to $z=1.5$, beyond which its contribution is negligible
(Appendix~\ref{sec:wings});}
(iii) the Gaussian covariance, which omits non-Gaussian contributions; and
(iv) the MTNG-like gas transfer function used to paint the electron
field.
None of these affects the estimator's construction or its validated
theoretical description, and \rev{we estimate that they} affect the accuracy of the forecast numbers at the \rev{$\sim20\%$} level.

Future work will include:
(a) a measurement with DESI LRG velocity reconstructions and Simons Observatory
temperature maps combined with early lensing data from LSST; and
(b) a joint shear--kSZ + cosmic shear analysis using a directly constrained matter
power suppression via Eq.~\eqref{eq:suppression_exact}.

\appendix

\section{Edge-bin contributions from beyond the tracer volume}
\label{app:edges}

The calibration matrix $A_{ij}$ of Eq.~\eqref{eq:Aij_def} is measurable only
where tracers exist, but the kSZ integral extends over the full line of sight.
The outermost bins therefore receive an uncalibrated contribution
\begin{equation}
\Delta C_\ell^{TV_i} =
\frac{T_{\rm CMB}}{c^2}
\int_{\rm outside} d\chi\;
\taup(\chi)\,\xi^v_i(\chi)\,
W_\kappa(\chi)\,\frac{\Pme(k_\ell;z)}{\chi^2},
\label{eq:edge_leakage}
\end{equation}
where the integral runs over the kSZ range outside the tracer volume and
$\xi^v_i(\chi)$ can be evaluated from the bin-averaged linear-theory
correlation, Eqs.~\eqref{eq:xiv_structure} and \eqref{eq:Psi_par}, normalized
to the measured $A_{ii}$ at zero lag.
Because the excluded region lies within the positive correlation core of the
outermost templates, the contribution is positive and decays inward over
roughly one correlation length, in agreement with the excess measured in
Fig.~\ref{fig:residual_scan}.
On real data, where the electron distribution beyond the reconstruction volume
is not directly measurable, Eq.~\eqref{eq:edge_leakage} supplies the
correction template (with the velocity correlation from linear theory and the
optical depth from the same model used for the interior bins), or the affected
bins may simply be excluded at a cost of
$\sim2/N_{\rm bin}\approx8\%$ of the statistical power per survey edge.
\rev{The same integral, extended over the full line of sight, gives the
negative-tail contribution of the gas outside the tracer volume to {all}
shells (Appendix~\ref{sec:wings}); the edge excess discussed here is the
positive-core part of that integral, which dominates only for the outermost
one to two shells. We also note that the simulated kSZ map extends one
correlation length ($\simeq115\;\mpc$) below the first shell, which is the origin
of the positive excess of bin~0 in Fig.~\ref{fig:residual_scan}.}

\section{Pipeline consistency tests}
\label{app:pipeline}

\subsection{Independence of noise realisations}

The bias check of Section~\ref{sec:bias_check} requires that the primary CMB
realisation and the LSST shape-noise realisation be statistically independent.
During this analysis we found that generating both Gaussian fields from the
same random seed induces a spurious cross-correlation
$C_\ell^{T_{\rm CMB}\kappa_{\rm noise}} =
\sqrt{C_\ell^{TT}\,N_\ell^{\kappa}}$ at all multipoles: the two
harmonic coefficient sets are then identical up to their amplitude
spectra, which biases $\hat{C}_\ell^{TV}$ by up to two orders of magnitude
above the signal.
After assigning distinct seeds, the fraction of multipoles with positive
$\hat{C}_\ell^{TV}$ dropped from $89.7\%$ to $51.8\%$, consistent with the
statistical expectation ($50\%$) for an unbiased estimator.
We flag this because the same failure mode arises naturally in forecast
pipelines that recycle a random generator across mock components.

\subsection{Scale dependence of the measurement uncertainty}

Figure~\ref{fig:sigma_vs_ell} shows the per-band Gaussian uncertainty
$\sigma[D_\ell^{TV_i}]$ of Eq.~\eqref{eq:per_bin_sigma}, band-averaged in the
logarithmic $\ell$ bins used throughout, for the five representative bins in
the two ideal configurations. Two regimes are visible.
At low multipoles the error is dominated by mode counting: a logarithmic band
centred at $\ell$ contains $\sim\ell^2\fsky$ modes, driving the steep initial
decline.
Through $\ell\sim10^2$--$3\times10^3$ the declining mode-counting factor is
partially compensated by the shape of $C_\ell^{TT}C_\ell^{V_iV_i}$, producing
the broad plateau, before the falling $V_i$ spectrum steepens the decline
again at the highest multipoles.
The bin-to-bin ordering follows the template variance
$\langle\tilde v_i^2\rangle$; in the LSST panel the plateau is higher and
flatter because shape noise replaces $C_\ell^{\kappa\kappa}$ as the dominant
term inside $C_\ell^{V_iV_i}$.
Read against the signal spectra of Fig.~\ref{fig:bins_comparison}, which peak
at $\ell\sim2$--$4\times10^3$ while $\sigma[D_\ell^{TV_i}]$ is still
declining, this figure locates the per-band SNR maximum in the same range and
makes explicit why individual bands are detected in the ideal configurations
while the observed cases rely on the Fisher combination over bins and
multipoles.

\subsection{Measurement and model comparison}

\begin{figure}[t]
\centering
\includegraphics[width=\linewidth]{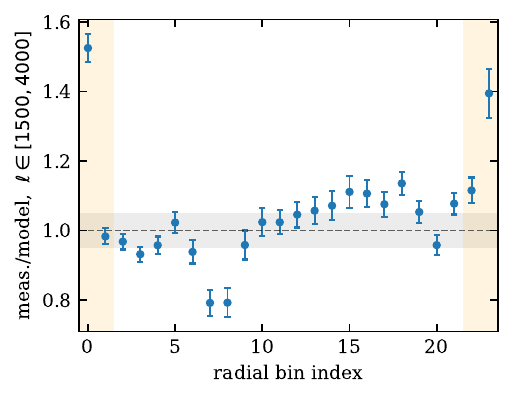}
\caption{
Measurement-to-model ratio in the peak band $\ell\in[1500,4000]$ for all 24
radial bins (points with Gaussian errors), summarizing the validation.
On average, the interior bins agree at the $\lesssim5\%$ (median) level. 
The shaded outer bins exhibit the positive excess from kSZ contributions
beyond the tracer volume (Appendix~\ref{app:edges}). The fluctuations in the points around one are the result of the long-wavelength velocity modulation.
}
\label{fig:residual_scan}
\end{figure}

In this section, we study the comparison between theory and model across all 24 bins by condensing it to
one number per bin: the measurement-to-model ratio in the peak band
$\ell\in[1500,4000]$.
Over the interior bins, excluding the two anomalous groups discussed next, the
parameter-free model matches the measurement with a median accuracy of
$\lesssim 5\%$ over the signal-dominated range $\ell\gtrsim10^3$.
At $\ell\lesssim500$ the measurement rises above the prediction, as expected
from the neglected Wick contraction and the convolution floor
(Section~\ref{sec:theory_validity}). The remaining differences between theory and measurement are due to several simplifications in our modeling (e.g., adopting the halo velocity field as the second leg rather than the particle one, as well as sample variance in both the velocity as well as density fields).

Figure~\ref{fig:residual_scan} displays deviations from the theoretical prediction with two notable features,
which are understood as follows.
First, the outermost one to two bins on each side show a strong positive
excess ($\sim35$--$50\%$ at $b=0$ and $b=23$, decaying inward over roughly one
velocity correlation length).
This is leakage from beyond the tracer volume: the kSZ map contains
ionized gas slightly outside the radial range spanned by the tracers, whose
velocities lie within the positive correlation core of the outermost
templates, but which the calibration matrix, which is measurable only where tracers
exist, cannot account for.
The effect is not a simulation artefact; on real data the kSZ integrates the
entire line of sight, so the outermost reconstruction bins always receive such
contributions.
They can either be modelled with the linear-theory correlation of
Eq.~\eqref{eq:Psi_par} extended beyond the survey (Appendix~\ref{app:edges})
or, as we do for the quoted accuracy, excised.
Second, we observe coherent fluctuations around one, as large as $\sim20\%$ for bins 7 and 8. Their origin is the sample modulation of the long-wavelength radial velocity modes.
The simulation-calibrated velocity kernel, measured from the same  realization, removes most but not all of this modulation.
The residual reflects the difference between the halo-sampled calibration
velocities and the mass-weighted velocities entering the kSZ, together with
the per-bin density-leg realization. In the limit of a large foreground
sample volume, such effects become negligible.

\section{Recovering the matter power suppression: a proof of concept}
\label{app:suppression}

\begin{figure}[t]
\centering
\includegraphics[width=\columnwidth]{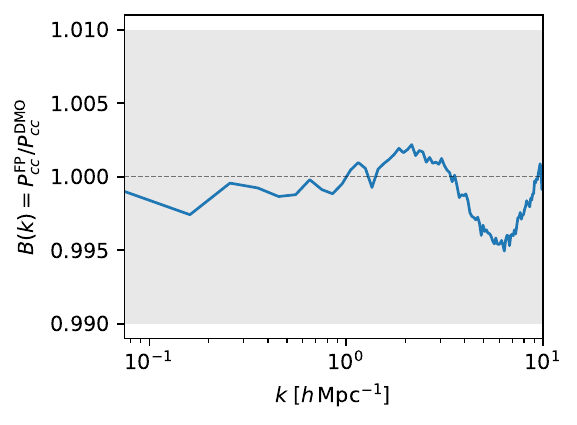}
\caption{
Dark-matter backreaction in MillenniumTNG at $z=0.5$: the ratio
$B(k) = P_{cc}^{\rm FP}/P_{cc}^{\rm DMO}$ of the cold-dark-matter auto-spectrum
in the full-physics run to that in the dark-matter-only run.
The deviation from unity (dashed line; shaded band $\pm1\%$) quantifies how
baryonic feedback alters the dark-matter clustering itself, and enters the
suppression inference only through the reference denominator
$P_{mm}^{\rm DMO}=P_{cc}^{\rm DMO}$.
It remains below the percent level across the scales relevant to cosmic shear.
}
\label{fig:backreaction}
\end{figure}

\begin{figure}[t]
\centering
\includegraphics[width=\columnwidth]{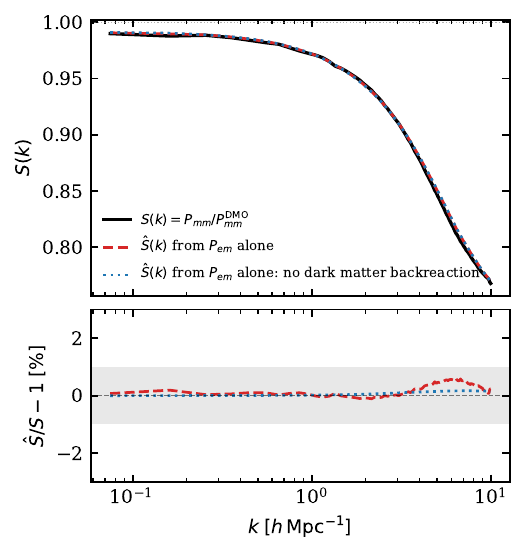}
\caption{
Proof-of-concept test of the suppression inference on MillenniumTNG at $z=0.5$.
\emph{Top}: the true matter power suppression $S(k)$ (solid black) compared
with the estimate $\hat S(k)$ reconstructed from the electron--matter
cross-spectrum $P_{me}$ alone through the species identity of
Eq.~\eqref{eq:suppression_exact}, assuming a known stellar fraction
($f_\star=1.5\%$) and $r_{me}=1$.
The reconstruction is shown both including the dark-matter backreaction
correction of Fig.~\ref{fig:backreaction} (red dashed) and neglecting it
entirely (blue dotted).
\emph{Bottom}: fractional deviation from the truth, each estimate compared to
the true suppression built with the same reference denominator (shaded band
$\pm1\%$).
The closure itself is accurate to $<0.2\%$; neglecting the backreaction of the
dark-matter-only reference degrades this only to $<0.5\%$, so the suppression
is recovered from $P_{me}$ to better than $0.5\%$ in either case.}
\label{fig:suppression_closure}
\end{figure}

The identity of Eq.~\eqref{eq:suppression_exact} converts the electron--matter
cross-spectrum $P_{me}$, which is the quantity we measure, into the matter power
suppression $S(k)=P_{mm}/P_{mm}^{\rm DMO}$.
Here we demonstrate the full chain on the MillenniumTNG hydrodynamical
simulation at $z=0.5$, as a proof of concept.

We start from $P_{me}$ and infer the baryon--matter cross-spectrum $P_{mb}$
assuming a known stellar mass fraction, fixed to $f_\star=1.5\%$.
This step is essentially exact for our purposes: stars form on scales
$\lesssim100\;h^{-1}{\rm kpc}$, well below the range $k\lesssim10\,h\,{\rm
Mpc}^{-1}$ relevant to cosmic shear, so on these scales the electrons trace the
baryons up to the constant $f_\star$ and $P_{mb}$ is recovered from $P_{me}$
without loss.
We then apply Eq.~\eqref{eq:suppression_exact}
with the closure assumption $r_{me}=1$, whose accuracy we assess directly below.
The dominant terms in the reconstruction are $f_c^2 P_{cc}+2f_b P_{mb}$, so the
$r_{me}=1$ approximation, which enters only through the small
$f_b^2 P_{bb}$ term, is expected to be subdominant.

The reconstruction requires a dark-matter-only reference in the denominator,
$P_{mm}^{\rm DMO}=P_{cc}^{\rm DMO}$ (for $f_\star=0$ in the reference).
Baryonic feedback modifies the dark-matter clustering itself, so
$P_{cc}^{\rm DMO}$ differs from the full-physics cold-dark-matter auto-spectrum
$P_{cc}^{\rm FP}$ that we measure; Figure~\ref{fig:backreaction} shows this
backreaction ratio $B(k)=P_{cc}^{\rm FP}/P_{cc}^{\rm DMO}$, which stays below
one percent on the scales of interest.
In an analysis of real data, neither $B(k)$ nor $f_\star$ is measured
internally; but both can be supplied by external models
\citep[e.g.,][]{vanDaalen2020, Mead2021}.
We therefore bracket the outcome by performing the reconstruction both with the
measured backreaction correction applied and with it neglected entirely.

Figure~\ref{fig:suppression_closure} shows the result.
The reconstructed suppression tracks the truth to better than $0.2\%$ when the
reference is treated self-consistently, confirming that the $r_{me}=1$ closure
and the species algebra are an excellent approximation across the full range of
scales. Neglecting the dark-matter backreaction of the reference degrades the
recovery only to $0.5\%$.
The suppression is thus reconstructed from $P_{me}$ to better than half a percent
whether or not an external backreaction model is supplied, and an external
model (or a matched dark-matter-only run) would remove even this residual.

\section{Error propagation to matter suppression}
\label{app:suppression_propagation}

\begin{table}[t]
\centering
\begin{tabular}{cccccc}
\hline\hline
$x \equiv P_{mb}/P_{mm}^{\rm DMO}$ & $(1-x)/\sigma_x$ & $S(x)$ & \ \ \ $\sigma_S$ \ \ \ & $S/\sigma_S$ & $({1-S})/{\sigma_S}$ \\
\hline
0.1 & 144.00 & 0.741 & 0.002 & 384.89 & 134.4 \\
0.2 & 64.00 & 0.772 & 0.004 & 204.21 & 60.4 \\
0.3 & 37.33 & 0.802 & 0.006 & 143.88 & 35.6 \\
0.4 & 24.00 & 0.831 & 0.007 & 113.66 & 23.1 \\
0.5 & 16.00 & 0.860 & 0.009 & 95.48 & 15.5 \\
0.6 & 10.67 & 0.889 & 0.011 & 83.33 & 10.4 \\
0.7 & 6.86 & 0.917 & 0.012 & 74.62 & 6.7 \\
0.8 & 4.00 & 0.945 & 0.014 & 68.07 & 4.0 \\
0.9 & 1.78 & 0.973 & 0.015 & 62.95 & 1.8 \\
1.0 & 0.00 & 1.000 & 0.017 & 58.85 & 0.0 \\
\hline\hline
\end{tabular}
\caption{
Propagation of a fixed-significance measurement of the baryon--matter
cross-spectrum into the inferred matter power suppression, at fixed $k$ and
cosmology and a $16\sigma$ detection of $P_{bm}$ \rev{at that $k$}
($\sigma_x/x = 6.25\%$), for $f_b = \Omega_b/\Omega_m = 0.157$ and a known
stellar mass fraction\rev{; see Table~\ref{tab:kbins} for the per-$k$
significances actually reached by the forecast}.
The measured ratio $x \equiv P_{bm}/P_{mm}^{\rm DMO}$ parametrizes the overall
suppression; $S(x)$ is the inferred suppression from the self-consistent
$r_{bm}=1$ closure, $\sigma_S = (dS/dx)\,\sigma_x$ its propagated uncertainty,
and $(1-S)/\sigma_S$ the detection significance of baryonic feedback itself.
Because $S$ responds to the cross-spectrum only through the baryon-fraction
weight $2f_b$, the fractional error is compressed by a factor
$\sim1/f_b$ relative to $P_{bm}$: across most of the range the feedback
suppression is determined to better than $1\%$ and detected at
$\gtrsim10\sigma$.}
\label{tab:propagation}
\end{table}

The identity of Eq.~\eqref{eq:suppression_exact} converts a measurement of the
baryon--matter cross-spectrum into the matter power suppression
$S(k) = P_{mm}/P_{mm}^{\rm DMO}$.
Here we quantify how the measurement precision propagates\rev{, both for a
generic fixed-$k$ significance and for the actual per-$k$ constraints of the
forecast of Section~\ref{sec:snr_results}}.

\rev{The total significances of Table~\ref{tab:snr} combine all multipoles and
shells and correspond to the precision on an overall amplitude of the signal
at fixed scale and redshift dependence; they do not translate directly into a
constraint on $\Pme$ at each $k$. To obtain the latter we repeat the Fisher
analysis with the amplitude of $\Pme$ in each of five bins of
$k_\ell=(\ell+1/2)/\chi_i$ as free parameters, using the full
shell--shell covariance of Eq.~\eqref{eq:covariance} and marginalizing over the
other bins. Table~\ref{tab:kbins} lists the result for the baseline
configuration (case 2b), the photometric full-footprint configuration (case 2e)
and the best LSST-alone configurations with SO and Advanced SO (cases 2h and
2i). The information is concentrated at $k\simeq1$--$4\,h\,{\rm Mpc}^{-1}$,
where the baseline cases measure the cross-spectrum amplitude to $17$--$34\%$
per bin, i.e.\ $S(k)$ to $3$--$7\%$ for a representative suppression, and the
best cases to $8$--$22\%$, i.e.\ $S(k)$ to $1.6$--$4\%$; the larger tracer volume
of the latter also opens the $4<k<8\,h\,{\rm Mpc}^{-1}$ bin, because the
lower-redshift shells map a given multipole to a higher $k$. This is the level at
which the forecast constrains the scale dependence of the baryonic
suppression; the generic propagation of Table~\ref{tab:propagation} below,
which assumes a given fixed-$k$ significance, should be read with the per-bin
significances of Table~\ref{tab:kbins} in mind rather than with the total of
Table~\ref{tab:snr}.}

\begin{table*}[t]\color{black}
\centering
\begin{tabular}{l|cc|cc|cc|cc}
\hline\hline
$k$ bin [$h\,{\rm Mpc}^{-1}$] & \multicolumn{2}{c}{case 2b} & \multicolumn{2}{c}{case 2e} & \multicolumn{2}{c}{case 2h} & \multicolumn{2}{c}{case 2i} \\
 & $\sigma_x/x$ & $\sigma_S/S$ & $\sigma_x/x$ & $\sigma_S/S$ & $\sigma_x/x$ & $\sigma_S/S$ & $\sigma_x/x$ & $\sigma_S/S$ \\
\hline
$[0.3,0.6)$ & -- & -- & -- & -- & -- & -- & -- & -- \\
$[0.6,1.0)$ & -- & -- & -- & -- & -- & -- & -- & -- \\
$[1.0,2.0)$ & $27\%$ & $5.2\%$ & $34\%$ & $6.5\%$ & $22\%$ & $4.2\%$ & $19\%$ & $3.6\%$ \\
$[2.0,4.0)$ & $17\%$ & $3.3\%$ & $21\%$ & $4.0\%$ & $11\%$ & $2.0\%$ & $8\%$ & $1.6\%$ \\
$[4.0,8.0)$ & $90\%$ & $17.3\%$ & $109\%$ & $20.9\%$ & $19\%$ & $3.6\%$ & $14\%$ & $2.7\%$ \\
\hline\hline
\end{tabular}
\caption{Marginalized fractional uncertainty on the amplitude $x$ of the
electron--matter cross-spectrum in bins of $k_\ell=(\ell+1/2)/\chi_i$, from the
full shell--multipole Fisher matrix of the forecast, and the corresponding
uncertainty on the suppression $S$ for a representative $x=0.6$ ($S=0.89$) from
Eq.~\eqref{eq:suppression_exact}. Columns: the baseline configuration (case 2b,
$\fsky=0.2$), the photometric full-footprint configuration (case 2e,
$\fsky=0.45$), and the best LSST-alone configuration with SO and with Advanced
SO (cases 2h and 2i, $\fsky=0.45$) of Table~\ref{tab:snr}. A dash denotes an
unconstrained bin ($\sigma_x/x>1$). The sum in quadrature of the unmarginalized
per-bin significances reproduces the totals of Table~\ref{tab:snr} to within
$10\%$.}
\label{tab:kbins}
\end{table*}

In this appendix, we work at fixed $k$ and cosmology and assume the stellar mass fraction is
externally known, so that the free-electron and baryon fields are related by a
fixed factor; under this assumption a \rev{given-significance} measurement of the
electron--matter cross-spectrum $P_{me}$, which shear--kSZ measures, is
equivalently a measurement of $P_{bm}$ \rev{at the same significance}, and we phrase the
propagation in terms of the latter.
Let $x \equiv P_{bm}/P_{mm}^{\rm DMO}$ be the measured ratio, which ranges
from $0$ (all baryons expelled) to $1$ (baryons trace the dark matter) and
thus serves as a proxy for the overall strength of the suppression.
Solving the closure of Eq.~\eqref{eq:suppression_exact} with $r_{bm}=1$
self-consistently gives $S(x)$, and linear propagation gives
$\sigma_S = (dS/dx)\,\sigma_x$ with
$dS/dx = 2f_b(S-f_b x)/(2S - f_c^2 - 2f_b x)$.
We adopt $f_b = \Omega_b/\Omega_m = 0.157$
and fix the fractional precision on the cross-spectrum to
$\sigma_x/x = 1/16 = 6.25\%$\rev{, a value that illustrates the compression of the error in going from
$P_{bm}$ to $S$; Table~\ref{tab:kbins} gives the per-bin significances that the
forecast actually reaches}.

The SNR-dominant multipole range, $\ell \sim 1000$--$8000$
(Section~\ref{sec:snr_results}), corresponds at the effective redshift of our
DESI-like LRG sample ($z_{\rm eff}\approx0.75$, $\chi\approx1840\;\mpc$) to
$k \approx 0.5$--$4\,h\,{\rm Mpc}^{-1}$ in the Limber approximation.
This is precisely the regime, in which
baryonic feedback suppresses the matter power spectrum and in which that
suppression is the dominant theoretical systematic for Stage-IV cosmic-shear
analyses.

Table~\ref{tab:propagation} tabulates the result.
The central feature is that $S$ depends on the cross-spectrum only through the
baryon-fraction weight $2f_b$, so fractional errors compress by roughly a
factor of $1/f_b \approx 6$ (more precisely, $d\ln S/d\ln x \approx 0.25$) in
propagating from $P_{bm}$ to $S$: a $6\%$ measurement of the cross-spectrum
determines the suppression to $0.3$--$1.7\%$ across the full range of $x$.
For a representative $40\%$ cross-spectrum suppression ($x=0.6$), the true
suppression is $S = 0.89$, measured to
$\sigma_S/S = 1.2\%$ and detected at $10\sigma$; for the stronger suppressions
$x\lesssim0.5$ the significance of feedback exceeds $15\sigma$, and for
several cases $(1-S)/\sigma_S > 100$. Thus with even a detection of 
10--20$\sigma$ of the electron--matter cross spectrum \rev{in a given $k$ bin}, one can measure the 
feedback suppression \rev{at that scale} to better than $1\%$ in many scenarios\rev{; for the
per-bin significances of Table~\ref{tab:kbins}, $S(k)$ is determined to
$3$--$7\%$ at $k\simeq1$--$4\,h\,{\rm Mpc}^{-1}$ in the baseline forecast and to
$1.6$--$4\%$ in the best configurations}.
This is exactly the precision required to remove baryonic feedback as a
systematic in weak-lensing cosmology, achieved here from a cross-correlation
that is immune to the galaxy-bias and intrinsic-alignment uncertainties of
the shear signal itself.

We note finally that recent measurements favor stronger feedback than
predicted by many hydrodynamical simulations
\citep[e.g.,][]{2025PhRvD.112h3509H, 2025MNRAS.540..143M, 2024MNRAS.534..655B, 2025PhRvD.112l3507H},
i.e.\ smaller $S$ and smaller $x$. If this is the case across mass, 
at fixed detection significance of $x$, the detection significance $(1-S)/\sigma_S$
grows as the suppression increases (Table~\ref{tab:propagation}).

\begin{revblock}\section{Contribution of the ionized gas outside the tracer volume}
\label{sec:wings}

\begin{figure*}[t]
\centering
\includegraphics[width=0.95\linewidth]{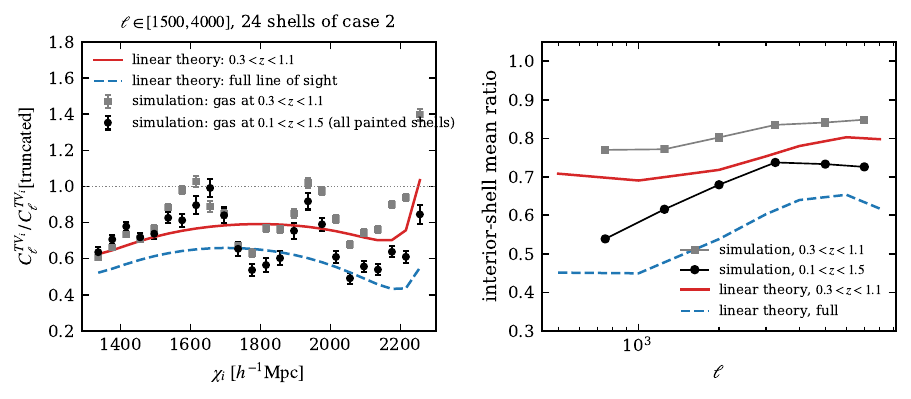}
\caption{
Effect of the ionized gas outside the tracer volume on the shear--kSZ signal.
\emph{Left}: ratio of $\hat{C}_\ell^{TV_i}$ measured with the kSZ map extended
to $0.3<z<1.1$ (grey squares) or to the full painted line of sight,
$0.1<z<1.5$ (black circles), to the value measured with the map truncated to
the tracer volume ($0.45<z<1$), averaged over $\ell\in[1500,4000]$ for the 24
velocity shells of case 2 (Gaussian errors) and compared with the linear-theory
prediction of Eq.~\eqref{eq:master_exact} (red; blue dashed). Because of the
negative tail of the line-of-sight velocity correlation, the added gas
\emph{reduces} the signal of the interior shells; the long-wavelength radial
modes of the realization modulate the ratio from shell to shell. \emph{Right}:
interior-shell mean of the ratio versus multipole. The suppression is stronger
at low $\ell$, where the density leg varies more slowly along the line of sight
and the cancellation set by the sum rule is more complete.}
\label{fig:wings}
\end{figure*}

Our simulated kSZ maps (Section~\ref{sec:ksz_map}) contain only the electrons
in the light-cone shells that span the tracer volume, in practice $0.45<z<1.0$,
so the model of Eq.~\eqref{eq:master_discrete}, with $A_{ij}$ calibrated over
the same volume, matches the measurement (Fig.~\ref{fig:theory_validation}) by
construction. The real kSZ map, however, integrates the whole line of sight.
Because the convergence is a broad projection, gas far outside the tracer
volume stays correlated with the template of shell $i$ through the long
negative tail of $\Psi_\parallel$, and the sum rule of Eq.~\eqref{eq:sumrule}
makes the size of the effect hard to guess: if the line-of-sight weight
$\taup W_\kappa\Pme/\chi^2$ in Eq.~\eqref{eq:master_exact} were constant, the
full-line-of-sight signal would vanish. What survives depends on how much this
weight varies across the support of the velocity correlation, about
$10^3\;\mpc$.

We quantify the effect in two ways: analytically, by evaluating
Eq.~\eqref{eq:master_exact} in linear theory with the line of sight integrated
over the range of the simulated map or over $0<z<2.3$; and directly, by
painting the kSZ of the remaining light-cone shells with the procedure of
Section~\ref{sec:ksz_map}, so that the simulated map contains all the ionized
gas at $0.1<z<1.5$, beyond which the lensing kernel and the velocity
correlation make the contribution negligible. Figure~\ref{fig:wings} compares
the cross-spectra of the same 24 templates with the truncated and the full
maps. The gas outside the tracer volume lowers the signal of the interior
shells by about $30\%$, more at low $\ell$, with a strong shell-to-shell
modulation by the long-wavelength radial modes of the realization. Most of the
loss comes from gas within $\sim500\;\mpc$ of the volume. Linear theory
reproduces this near part of the tail but overpredicts the far part, a
difference we attribute to the sample variance of the largest radial modes in a
single realization.

\begin{figure}[t]
\centering
\includegraphics[width=\linewidth]{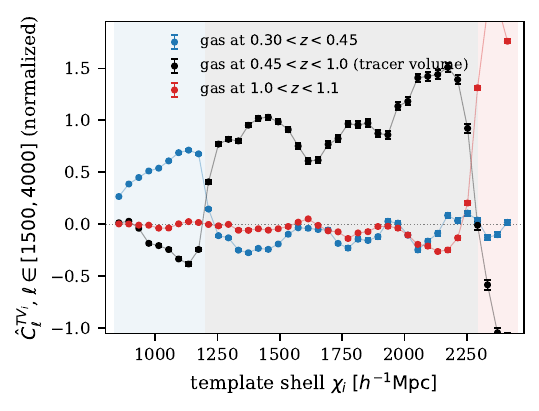}
\caption{
Direct test of the sign of the outside-volume contribution. Each of the three
simulated kSZ light-cone pieces (gas at $0.30<z<0.45$, blue; $0.45<z<1.0$, the
tracer volume, black; $1.0<z<1.1$, red; shaded bands) is correlated separately
with the 40 velocity templates of the extended volume $0.3<z<1.1$ (smoothed
true halo velocities, high-redshift source sample), averaged over
$\ell\in[1500,4000]$ and normalized to the mean of the interior tracer-volume
shells. Each piece correlates positively with the shells inside it and
\emph{negatively}, at $10$--$25\sigma$, with the shells more than $\sim50\;\mpc$
outside it: this is the negative tail of the line-of-sight velocity
correlation seen through the kSZ itself, and on the real sky every shell has
such gas on both sides.}
\label{fig:wings_profile}
\end{figure}

Figure~\ref{fig:wings_profile} shows the sign of the effect directly: each kSZ
piece anti-correlates, at $10$--$25\sigma$, with the velocity shells more than
$\sim50\;\mpc$ outside it, and correlates positively with the shells inside it
and with those within one core width of its edges, the edge excess of
Appendix~\ref{app:edges}.

All forecasts in this paper therefore use templates measured against the full
map. For the smoothed-true-velocity templates, the SNR of case 2b drops by
$28\%$, as linear theory calibrated on the simulation predicts. The
reconstructed templates of Section~\ref{sec:recon} lose about half: the
reconstructed velocity field is a potential flow built from the smoothed galaxy
density, which obeys the same sum rule but has a deeper negative radial tail
than the true halo velocities, and when the reconstruction stops at the tracer
volume, the gas just outside it is also uncorrelated with the modes the
reconstruction can recover.

We note that this loss is a property of the estimator, not a bias: on real data
one measures or predicts $A_{ij}$ over the full range where the gas lives, so
the outside contribution is part of the model, at the price in significance
quantified here. For a single broad lensing kernel this price is unavoidable;
combining tomographic source bins into kernels of compact radial support, as in
the nulling approach of Ref.~\citep{2026arXiv260729091G}, reduces the
cancellation (Section~\ref{sec:extvol}). Finally, we stress that the
tomographic binning does not localize the kSZ signal to the tracer shell: the
binning sets the velocity kernel, but the non-zero response comes from the
incomplete cancellation of that kernel against the slowly varying line-of-sight
weights.
\end{revblock}

\begin{revblock}\section{Analytic forecast}
\label{app:analytic}

At multipoles $\ell\gg L_v$ the convolution of Eq.~\eqref{eq:convolution}
collapses and both the signal and the covariance of the estimator factorize
into a velocity part and a density part. For a set of templates
$V_{ia}=\vtil_i\kappa_a$ (shell $i$, source sample $a$),
\begin{align}
s_{ia}(\ell) &= -\frac{T_{\rm CMB}B_\ell}{c^2}\!\int\! d\chi\,\taup
W_a\,\frac{\Pme(k_\ell,z)}{\chi^2}\,\xi_i(\chi),\\
\mathrm{Cov}_{(ia)(jb)}(\ell) &= \frac{C_\ell^{TT}}{(2\ell+1)\fsky}\;
\frac{\Xi_{ij}}{c^2}\,\bigl[C_\ell^{\kappa_a\kappa_b}+N_a\delta_{ab}\bigr],
\end{align}
where $\xi_i(\chi)$ is the shell-averaged velocity correlation of
Eq.~\eqref{eq:xiv_structure}, smoothed on one leg, and $\Xi_{ij}$ the template
covariance, smoothed on both legs, both from the linear-theory $\Psi_\parallel$
of Eq.~\eqref{eq:Psi_par}; $C_\ell^{\kappa_a\kappa_b}$ are the Limber
cross-spectra of the source samples, $N_a$ their shape noise, $\Pme=T_{\rm
gas}P_{mm}$ as in Eq.~\eqref{eq:transfer}, and $\taup$ assumes fully ionized
hydrogen and helium. The signal-to-noise ratio follows from
Eq.~\eqref{eq:fisher_snr} over the allowed pairs $(i,a)$, with the beam,
$C_\ell^{TT}$, source distributions and shell widths of the simulation.

With physical constants and no free parameter, the forecast reproduces the
simulation to $15\%$ for case 2b with the kSZ restricted to the tracer volume,
and overpredicts the loss from the full line of sight, because linear theory
overpredicts the far tail of the velocity correlation
(Appendix~\ref{sec:wings}). We absorb both into a single normalization factor of
$1.07$, set by the measured full-line-of-sight value of case 2b, and apply it
to all entries of Table~\ref{tab:extvol}; the extended-volume simulations
(cases 2f--2i) then agree with the normalized forecast to $\lesssim10\%$. Shell
by shell, the analytic information per mode and the inter-shell correlations
agree with the simulation at the multipoles that carry the signal, and the
occupancy dilution of the simulated templates (Section~\ref{sec:vfield})
cancels in the ratio, as it should. The only approximations are the
linear-theory velocity correlation and the separable covariance, and the result
does not depend on the shell width, so the forecast is well suited to comparing
configurations; its absolute values carry the $\approx10\%$ uncertainty of the
normalization.
\end{revblock}

\acknowledgements

We are grateful to Simone Ferraro, Fiona McCarthy and Blake Sherwin
for insightful comments during the preparation 
of this manuscript. \rev{We thank Anoma Ganguly and Elisabeth Krause for
useful discussions, and the anonymous referee for a careful and constructive
report that led to substantial improvements of the forecast.}
This research used the CBorg AI platform and resources provided by the IT Division at the Lawrence Berkeley National Laboratory (Supported by the Director, Office of Science, Office of Basic Energy Sciences, of the U.S. Department of Energy under Contract No. DE-AC02-05CH11231). We thank the AbacusSummit team for making the simulation data publicly available.
This research used resources of the National Energy Research Scientific
Computing Center (NERSC), a US Department of Energy Office of Science User
Facility. 

A nameless poet dedicated the following proverb to this work: ``Baryons that keep their hair are exactly the ones kSZ can see -- a stripped, no-hair gas profile is what it cannot.''

\bibliographystyle{apsrev4-1}
\bibliography{main}

\end{document}